\documentclass{article}%
\pdfoutput=1
\usepackage{amsmath}
\usepackage{amsfonts}
\usepackage{amssymb}
\usepackage{graphicx}%

\usepackage{mathpazo}
\usepackage{hyperref}
\usepackage{multimedia}

\usepackage[space]{grffile}%

\providecommand{\U}[1]{\protect\rule{.1in}{.1in}}
\newtheorem{theorem}{Theorem}

\newtheorem{remark}[theorem]{Remark}

\begin{document}

\title{Wavelet Analysis of Big Data in the Global Investigation of Magnetic Field
Variations in Solar-Terrestrial Physics}
\author{Srebrov, Bozhidar\\\texttt{Institute of Mathematics and Informatics}\\\texttt{Bulgarian Academy of Sciences}
\and Kounchev, Ognyan\\\texttt{Institute of Mathematics and Informatics}\\\texttt{Bulgarian Academy of Sciences}
\and Simeonov, Georgi\\\texttt{Institute of Mathematics and Informatics}\\\texttt{Bulgarian Academy of Sciences} }
\maketitle

\begin{abstract}
We provide a Wavelet analysis of Big Data in Solar Terrestrial Physics. In
order to explain and predict the dynamics of the geomagnetic phenomena we
analyze high frequency time series data from different sources: 1. The
Interplanetary Magnetic Field (from the ACE satellite). 2. The Ionospheric
parameters - TEC (from ionospheric sounding stations). 3. The ground
Geomagnetic data (from ground geomagnetic observatories, located in middle
geographic latitudes).

We seek for correlations in the wavelet coefficients which explain the
dynamics of different magnetic phenomena in the Solar Terrestrial Physics.

The large variety of data used in our research from both Solar Astronomy and
Earth Observations makes it a contribution to the newly developing area of AstroGeoInformatics.

\end{abstract}

\section{Introduction to Big magnetic Data in Solar-Terrestrial Physics}

Following the present concepts, Big data in Solar-Terrestrial Physics would
refer to large amounts of measured data

\begin{enumerate}
\item whose source is not homogeneous

\item having considerable dimension

\item the size and the format excess the capacity of the conventional tools to
effectively capture, store, manage, analyze, and exploit them, and finally,

\item having a complex and dynamic relationship.
\end{enumerate}

Institutions are increasingly facing more and more Big data challenges, and a
wide variety of techniques have been developed and adapted to aggregate,
manipulate, organize, analyze, and visualize them. The techniques currently
applied in Big Astronomical and Earth Observation data usually draw from
several fields, including statistics, applied mathematics and computer
science, and institutions that intend to derive value from the data should
adopt a flexible, reliable, and multidisciplinary approach. In particular,
utilizing Big data in Solar-Terrestrial Physics and its analytics will improve
the performance of prediction mechanisms for unusual geomagnetic events as for
example geomagnetic storms.

Normally, Big data in Solar-Terrestrial Physics are accessible, but there are
fewer tools to get value out of them as the data are immediately available
only in their most coarsest form or in a semi-structured or unstructured format.

One broad way of using Big magnetic data in Solar-Terrestrial Physics to
unlock significant value is to collect the data at a tick-by-tick level, i.e.
at higher frequency. Let us remark that the standard registrations of the
geomagnetic field variations has the order of about $0.1-10$ mHz (i.e. of
periods $1,66$ min till $2,77$ hours). When considering geomagnetic data at
higher frequency (here, high frequency is understood from the point of view of
the standards in Geomagnetism), usually such data illustrate the complex
structure of irregularities and roughness (i.e., multifractal phenomena) due
to huge amounts of microstructure noise. The non-homogeneity characterized by
multifractal phenomena is caused by a large number of instantaneous changes in
the geomagnetic field due to geomagnetic storms and various sources of noises
as, for example, the low frequency plasma instability modes. Therefore, mining
big geomagnetic data needs to intelligently extract information conveyed at
different frequencies. At the present moment the registration of geomagnetic
signals has the maximal frequency of seconds, however majority of the
geomagnetic or ionosound data are still collected at maximal frequency $4$ or
$5$ min. as we will see below.

With the classic assumption of Data Mining, data are generated by certain
unknown function representing signals plus random noise (see e.g. one of the
bibles of modern pattern recognition \cite{bishop} and the references
therein). Decomposing big magnetic data in Solar-Terrestrial Physics is
equivalent to extracting the systematic patterns (i.e., approximate the
unknown function) conveyed in the data from noise, which is the standard
approach of the classic signal processing theory brilliantly presented in the
famous handbook \cite{oppenheim}.

The situation in the modeling of Geomagnetic fields and in particular
geomagnetic storms falls in the framework of analyzing jump events in Big
Data, which has been recently thoroughly studied by various researchers. In
particular, it has been studied in the context of financial time series, in
the nice research of Sun and Meinl, \cite{Sun2012}, \cite{Sun2015}. They point
out, that a specific problem arises when the trend component exhibits
occasional jumps that are in contrast to the slow evolving long term trend and
one needs to apply appropriate tools to analyze these singularity events.

In Geomagnetism, jumps are often caused by some unexpected large Geomagnetic
storms or by predictable changes in the sectorial structure of the
Interplanetary magnetic field. Traditional linear denoising methods (e.g.,
moving average) usually fail to capture these jumps accurately as these linear
methods tend to blur them. On the other hand, nonlinear filters are not
appropriate to smooth out these jumps sufficiently, because the patterns
extracted by nonlinear filters are not stationary to present long run dynamic
information. The situation is rather similar to that in the case of financial
time series, as noted in \cite{Sun2012} and in the references therein.

The present Chapter puts the foundations of a research in the Global structure
of the magnetic phenomena in Solar-Terrestrial Physics, by systematically
applying \textbf{Wavelet Analysis (WA)} to the available Big Data. We use
\textbf{Continuous Wavelet Transform (CWT)} to analyze the structure and the
dynamics of Geomagnetic storms. The wavelet method has been shown to be one of
a number of multifractal spectrum computing methods and proven to be a
reliable in signal processing as established in the classical monograph
\cite{mallat}. It has proven to be very suitable for analyzing time series
analysis -- for example, smoothing, denoising, and jump detection very diverse
areas of science, finance, economics, see e.g., \cite{percival},
\cite{gencay}, \cite{hubbard}, \cite{MeinlSun}, \cite{kounchevBOOK}.

An important advantage of the wavelet method in analyzing magnetic phenomena
in Solar-Terrestrial Physics, where factors of different scales interfere, is
that it performs a Multiresolution analysi (MRA), in other words, it allows us
to analyze the data at different scales (each one associated with a particular
frequency passband) at the same time. In this way, wavelets enable us to
identify single events truncated in one frequency range as well as coherent
structures across different scales. Many recent studies have applied wavelet
methods in mining geophysical and geomagnetic data, some very recent
references are \cite{xu}, \cite{xu2011}, \cite{Klausner2016},
\cite{Klausner2016b}, \cite{Klausner2017}, \cite{Schnepf2016}.

Let us recall that there are in fact two versions of Wavelet Analysis:
Continuous Wavelet Transform (CWT) and the Discrete Wavelet Transform (DWT).
In both of them, there is a large variety of wavelet functions by which one
may perform the signal decomposition. A common approach in choosing the
wavelet function is to use the shortest wavelet filter that can provide
reasonable results, see e.g. \cite{percival}. The main challenge in performing
WA is how to determine the combination of wavelet function, level of
decomposition, and threshold rule to reach an optimal smoothness that
generally improves the performance of classic models after denoising the data.
We have provided at the end a short Appendix which explains briefly the
technical details of WA and our choice of wavelet function.

Another major advantage of Wavelet Analysis (either discrete or continuous),
which makes it well adapted to the purposes of Big Data is that, similar to
the classical Fourier analysis, there exist very fast algorithms for
processing large amounts of data, \cite{mallat}.

Due to space restriction, in the present Chapter we have limited ourselves
with just preliminary research of the correlations which exist among the data
in the frequency domain (the coefficients of the CWT of the time series).
Although the results obtained are very interesting and promising, we have not
provided a more detailed statistical analysis (as e.g. in \cite{Sun2012},
\cite{Sun2015}) which would uncover much deeper and interesting connections
between the different factors playing role in the Geomagnetic phenomena. We
leave such analysis for follow up research which is in progress.

The structure of the Chapter is as follows: in section \ref{SectionMechanism}
we recall the Big Picture of the Solar-Terrestrial Mechanism -- in quiet and
in stormy periods. This has to give an idea to the unexperienced reader about
the complexity of the manifestation of a geomagnetic storm and the necessity
to apply modern methods of Machine learning to Big Data, for a deeper
understanding of the phenomena. In section \ref{sectionComponents} we provide
a short description of the different components of the ground geomagnetic
field provided by Chapman's analysis -- the global index $D_{st}$ and local
disturbance index $DS$ which show how complicated the structure of the
geomagnetic field on Earth is. In section \ref{sectionDisturbances} we provide
basic information about two famous geomagnetic storms. In section
\ref{sectionDataSources} we provide basic information about the different
types of data used for analyzing the Big picture of the magnetic phenomena in
the Solar-Terrestrial Physics. In section \ref{SectionExperiments} we provide
the results of the application of CWT to the $3$ main types of data in the
form of Simultaneous Time Series (Interplanetary Magnetic Field data,
Ionospheric TEC data, Geomagnetic data), in some quiet days and in days of
Geomagnetic storms. For every experiment carried out, we provide some
empirical observations on the correlations between the CWT coefficients (the
frequency domain) for simultaneous time series. In a forthcoming research we
will apply the methods of Statistical Analysis for a more rigorous analysis of
these observed correlations between the different types of data.

The large variety of data used, from both Solar Astronomy and Earth
Observations, makes our research a contribution to the newly developing area
of \textbf{AstroGeoInformatics}.

\section{Mechanism of generating strong geomagnetic storms (long period
geomagnetic field variations)\label{SectionMechanism}}

The main purpose of the present study is to analyze different sources of data,
and to discover correlations between the (relatively) high frequency
geomagnetic variations (wave packages with short period about $0.1$ mHz till
$10$ mHz) which happen not only during Geomagnetic storms but also in more
quiet periods.

Before carrying out such an analysis we will provide the global context of the
Geomagnetic picture in the Solar-Terrestrial interactions.

\subsection{The Big picture of the Solar-Terrestrial Physics - quiet and
disturbed Geomagnetic phenomena}

In the present section we will outline the Big picture of the influence of the
solar activity on the Interplanetary Magnetic field, the Ionospheric
parameters, and the (ground) Geomagnetic field.

First of all, we speak about events happening in the Magnetosphere, i.e. in
the region of space surrounding Earth where the dominant magnetic field is the
magnetic field of Earth, rather than the magnetic field of interplanetary
space. The Magnetosphere is formed by the interaction of the solar wind with
Earth's magnetic field. The Earth's magnetic field is continually changing as
it is buffeted by the solar wind.

\begin{figure}
[!ptb]
\begin{center}
\includegraphics[
height=3.0477in,
width=2.4942in
]%
{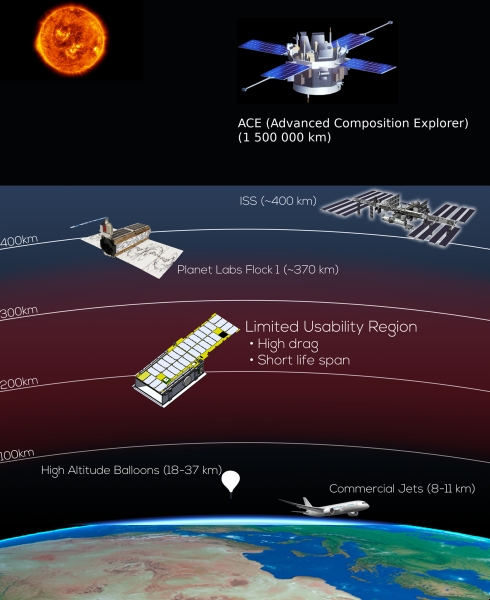}%
\caption{Scheme of the Interplanetary Space, Ionosphere and Earth, from which
we acquire data.}%
\label{Scheme_Ionosphere3}%
\end{center}
\end{figure}

\begin{enumerate}
\item In quiet periods the Sun emits a flow of particles (solar wind) having a
relatively constant intensity and speed, which start with appr. $50$ km/sec
and is accelerated to about appr. $360$ km/sec close to Earth (which is
obtained by the Parker model). The speed is accelerated by a mechanism which
is still not known but most probably due to energy transfer in the solar wind.

\item In a disturbed state, the hyper-activity of the Sun causes Coronal Mass
Ejection (CME) which increases the amount of charged particles; they have the
macro-speed about $1500-2500$ km/sec when they leave the Sun; this speed
decreases to about $500-700$ km/sec when they approach the Earth.

\item In the Interplanetary medium there is a collisionless plasma where the
particles are electrons and ions (protons); the \textbf{satellite ACE}
collects the data of the Interplanetary Magnetic Field (\textbf{IMF}) at the
distance $1.5$ Million km from Earth in this plasma (see Figure
\ref{Scheme_Ionosphere3}); the most important is the $Z-$component of IMF
called $B_{z}$ (which is perpendicular to the ecliptic) which influences the
formation of the storm; during the strong storms the component $B_{z}$ is negative.

\item After that, the flow of particles approaches the Magnetosphere (about
$12000-25000$ km) which is the belt of Van Allen (mathematical figure called
torus), where they are caught by the Earth's magnetic field (the Earth's
dipole); this looks like a cavern but they are mainly concentrated in the
equatorial domain. The so-called \textbf{Ring Current} is formed in the Van
Allen belts. In the Van Allen belts one does not have plasma, but there is a
kinetic movement of the different particles.

\item Then at about 1000 km from Earth they enter the Plasmosphere and the
Ionosphere which is a plasma (partially ionized gas), and the laws describing
the plasma state start to work. There are collisions among the particles.
Hence, more wave effects may arise compared to the collisionless plasma. The
ionosound stations acquire the values of the Ionospheric parameters at these
heights. For more details see the excellent exposition in the classical
monograph \cite{Mitchner}.

\item In the quiet days a regular source of disturbances in the earth's
ionosphere is also the solar terminator (at sunrise and sunset).
\end{enumerate}

On Figure \ref{Scheme_Ionosphere3} below we provide the overall picture of
different sources of measurements which we have used in our study.%

\begin{figure}
[!ptb]
\begin{center}
\includegraphics[
height=2.7275in,
width=3.8468in
]%
{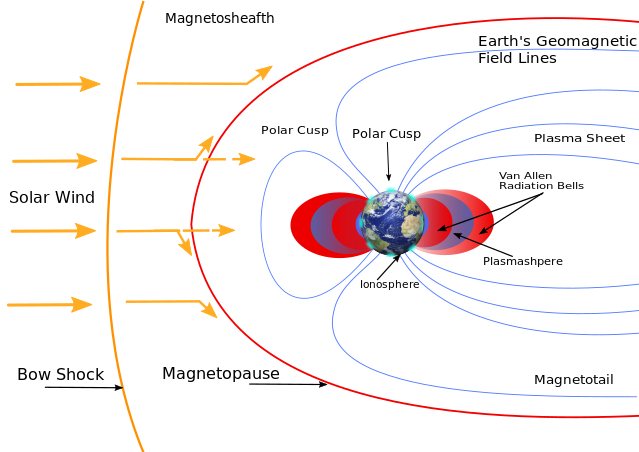}%
\caption{...}
\label{magnetic_field_drawing_m}%
\end{center}
\end{figure}

On Figure \ref{magnetic_field_drawing_m} we provide the Magnetic field structure.%

\subsection{The geomagnetic storms}

The geomagnetic storms are by their nature \textbf{long period geomagnetic
field variations}.

Geomagnetic storms are phenomena directly related to solar activity. They
result from the interaction of the magnetosphere and the ionosphere with
changes in the interplanetary conditions caused by closed interplanetary
magnetic structures. These structures are formed in the active centers of the
Sun's chromosome as a result of a sudden impulse ejection of the substance in
the quiet solar wind called Coronal Mass Ejection (CME).

The storm is mainly characterized by a decrease in the horizontal component
$H$ of the geomagnetic field, which encompasses the entire planet during a
geomagnetic storm. At low latitudes, for a long time, a current system called
the \textbf{Ring Current} is formed around the Earth at the distance
approximately $2$ to $4$ Earth radii. The nowadays idea of the size and
strength of the Ring current system gives us reason to view it as a toroid
inside the magnetosphere (in the area of radiation belts, called Van Allen
belts) and formed by particles of the solar wind. The geomagnetic field in the
magnetosphere captures particles from the interplanetary plasma by creating an
axisymmetric distribution of these particles in space. This leads to an
amplification of the Van Allen radiation belt. The perturbed geomagnetic field
associated with the so-called main storm phase in many cases is not
axisymmetric. The analysis of the perturbed field shows that it can be
represented as a sum of an axisymmetric part and an asymmetric component. This
indicates that the proton belt is substantially asymmetric, especially in the
early part of the main phase of the storm. Thus, asymmetry appears as an
essential feature of Ring Current formation.

The intensive \textbf{proton belt} during a storm significantly deforms the
magnetic field of the magnetosphere and changes its structure. In particular,
we see that the domain of particle trapping approaches the Earth, in the
interplanetary environment, and the polar ray oval shifts in the direction of
the equator.

The intensity of the storm strongly depends on the geographic latitude. Thus,
the decreasing of the horizontal component of the field during the storm in
the different ground magnetic observatories is similar in shape but different
in amplitude (see e.g. Figure \ref{FIG_v1_panag_storm1} with the $H-$component
of the geomagnetic field from observatory PAG in Bulgaria, and Figure
\ref{FIG_ds_sua_min_storm1__} with the $DS$ index of the geomagnetic field
data from observatory SUA in Romania). At low latitudes, this amplitude is
larger and decreases as the latitude increases. There is a difference of the
amplitude of the $H-$component of the magnetic field for the different
longitudes. This asymmetry of the field is related to the asymmetry of the
Ring Current considered above.

Geomagnetic storms are very diverse, but they are subdivided into two main
types - "standard" and "with sudden start". Geomagnetic storms of the second
type are characterized by the absence of a pronounced sudden start. But in
practice the main features of these storms during the main phase are like
those of the standard type. Therefore, the initial contraction of the
magnetosphere is not a prerequisite for the occurrence of the main phase of
the storm.

For our purposes it is enough to mention that the type of Sun activity (solar
chromospheric disturbances near or far from the solar equator) may cause
different types of storms. The formation and spread of the interplanetary
disturbed structure of plasma and the interplanetary magnetic field is a
complex magnetohydrodynamic process that is essentially three-dimensional. The
statistical analysis shows that more than $80\%$ of the strong geomagnetic
storms are associated with intensive processes in the active centers of the
Sun. For average storms, this percentage is smaller and is between $60-80$.
Briefly, the Solar astronomy may provide an important information which has to
be taken into account if predictions are needed.

It happens that important Sun processes \textbf{do not have any impact} on the
geomagnetic field. This can occur in a relatively small number of cases when
the disturbed interplanetary structure does not affect the point of the
Earth's orbit in which it is at that moment. The precise understanding of
these phenomena requires the joint efforts of astronomers and geophysicists.
All this indicates that knowing and predicting geomagnetic storms depends on
knowing the propagation of interplanetary disturbances as a hydrodynamic
process in three dimensions, \cite{srebrov2003}. For example, it has been
found that the direction of the interplanetary magnetic field (IMF) vector $B$
is essential to unlock the geomagnetic storm mechanism. In particular, if the
$Z-$component (denoted by $B_{z}$ in the coordinate system where the axis $Z$
is perpendicular to the ecliptic) is negative, then this indicates a
possibility for a very strong geomagnetic storm.

It would be interesting for the unexperienced reader to hear about the
parameters of a simulated \textbf{Coronal Mass Ejectio}n (CME), see details in
\cite{srebrov2003}, where a magnetohydrodynamic model is numerically
implemented: the release energy is E, = $6.0\times10^{22}$ J; the release mass
is M, = $2.5\times10^{10}$ kg; the initial velocities of the ejected flow are:
the radial $u_{d}=1500$ km / sec. and the tangential is $v_{d}=0$; the
duration of CME is $180$ seconds; the angle of the small conic area associated
with the CME is $27^{0}$, see a model with these realistic parameters
simulated in \cite{srebrov2003}. To understand how the southern direction of
the interplanetary magnetic field is formed, we will look at the results of
the computer modeling of the disturbance propagation in the interplanetary
environment caused by the CME. As more details are provided in
\cite{srebrov2003}, let us shortly describe the dynamics of the simulated
disturbance propagation in the interplanetary medium: After the CME happens at
$t=0$ and has duration $3$ min., about $40$ hours later the disturbance
reaches Earth's orbit and (see Figures \ref{fig_srebrov_1},
\ref{fig_srebrov_2} below and the Figures in \cite{srebrov2003}).

\begin{remark}
\textbf{REMOVE} = Remove:\ Figure \ref{fig_srebrov_1} and \ref{fig_srebrov_2})
the resulting tangential velocity $v$ of the disturbed solar wind is shown on
Figure \ref{fig_srebrov_1},%

\begin{figure}
[!ptb]
\begin{center}
\includegraphics[
height=3.1048in,
width=3.6778in
]%
{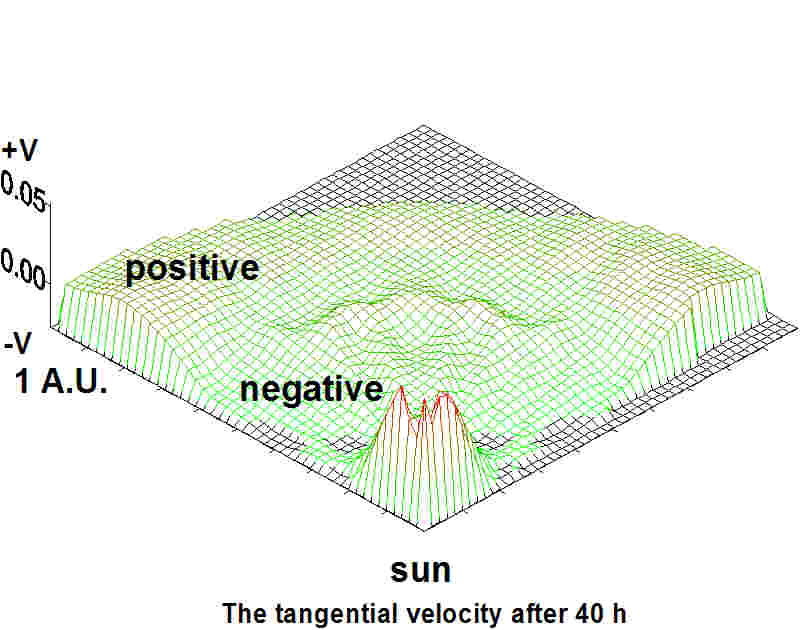}%
\caption{...}%
\label{fig_srebrov_1}%
\end{center}
\end{figure}

from which it is seen that $v$ has positive and negative values.
\end{remark}

\begin{remark}
\textbf{REMOVE = }This picture is in a meridional plane, defined by Sun-Earth
as $x$ axis, and $Z$ vertical to the ecliptic; in it the tangential velocity
$v$ is the component of the velocity along the $Z$ axis. The magnetic field
components are shown in Figure \ref{fig_srebrov_2}, which is again in the
meridional plane.%

\begin{figure}
[!ptb]
\begin{center}
\includegraphics[
height=2.2538in,
width=2.2609in
]%
{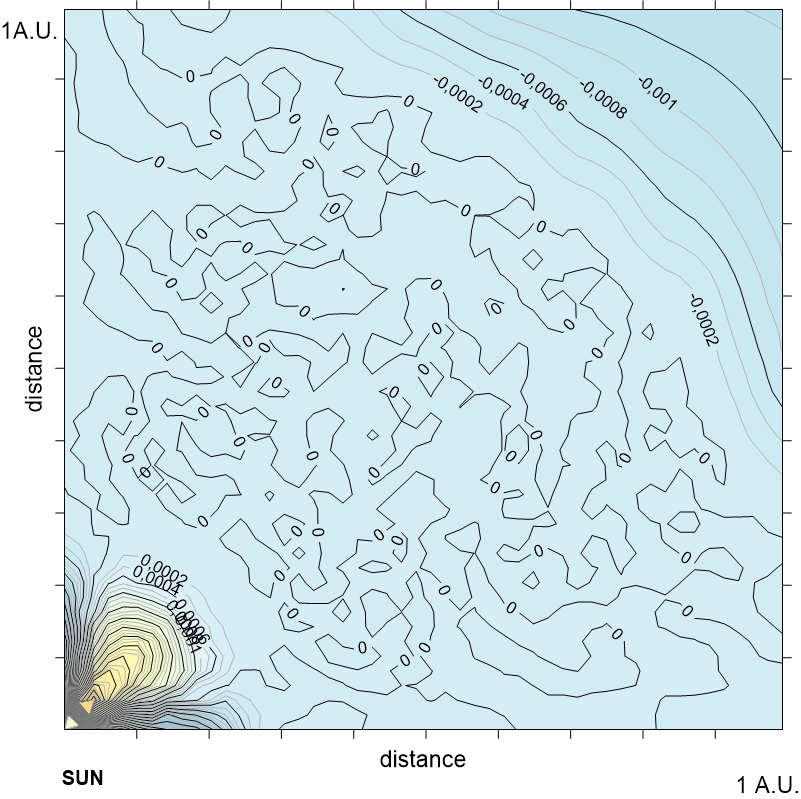}%
\caption{...}%
\label{fig_srebrov_2}%
\end{center}
\end{figure}

\end{remark}

\begin{figure}
[!ptb]
\begin{center}
\includegraphics[
height=1.9735in,
width=2.5545in
]%
{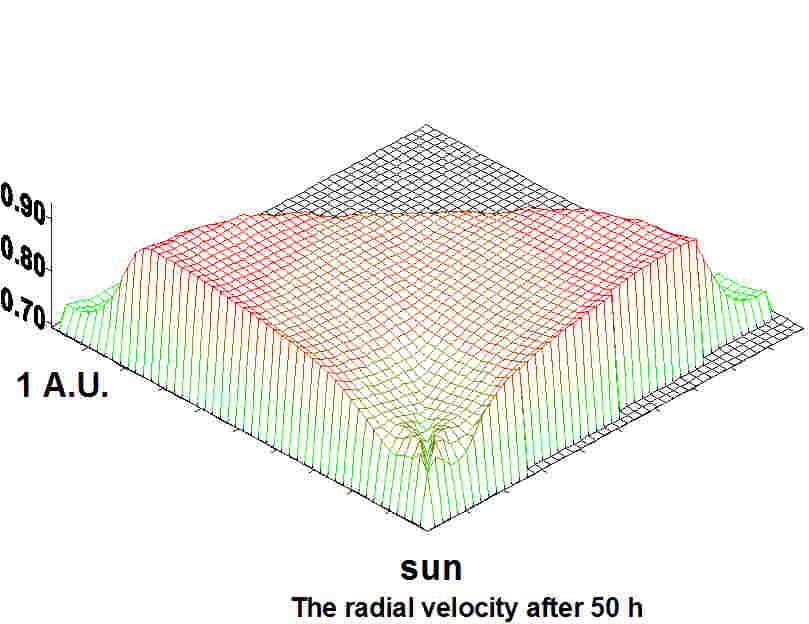}%
\caption{...}%
\label{Fig_unn50}%
\end{center}
\end{figure}

We put the following figures here: Below on Figure \ref{Fig_unn10},
\ref{Fig_unn30}, \ref{Fig_unn50} we provide the radial solar wind velocity $U$
at different times:\ $10,$ $30,$ $50$ hours after the CME; note that the value
of $U$ on the Figures is dimensionless:\ %

\begin{figure}
[!ptb]
\begin{center}
\includegraphics[
height=2.3159in,
width=2.8729in
]%
{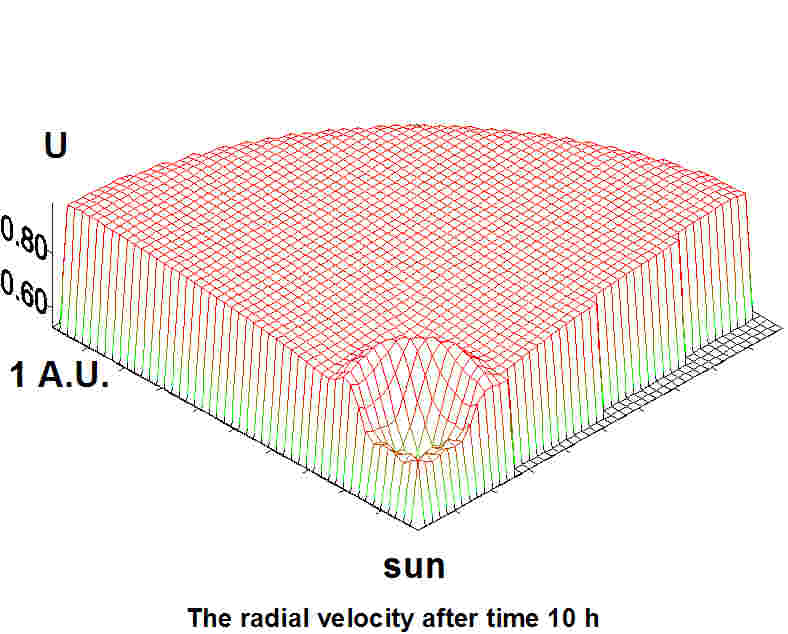}%
\caption{...}
\label{Fig_unn10}%
\end{center}
\end{figure}

%

\begin{figure}
[!ptb]
\begin{center}
\includegraphics[
height=2.0303in,
width=2.582in
]%
{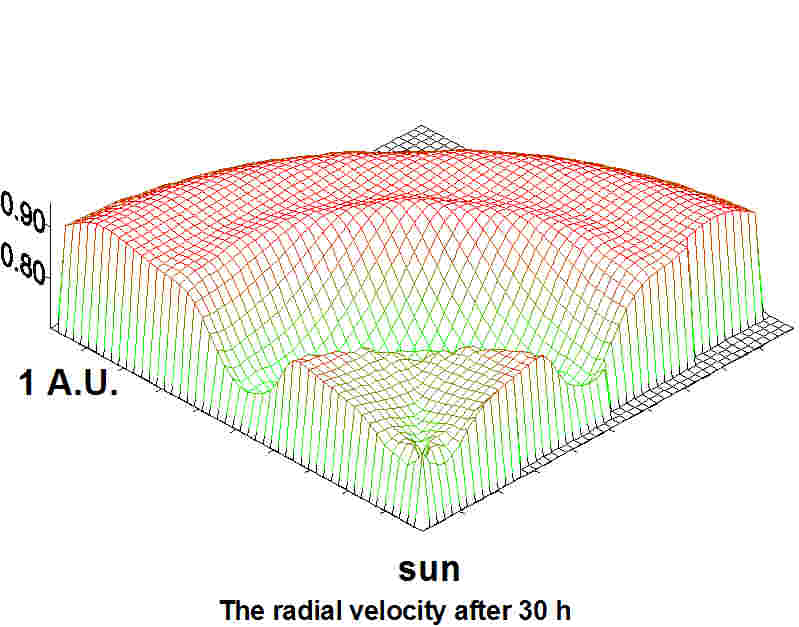}%
\caption{...}%
\label{Fig_unn30}%
\end{center}
\end{figure}

%

\begin{figure}
[!ptb]
\begin{center}
\includegraphics[
height=2.2671in,
width=2.9111in
]%
{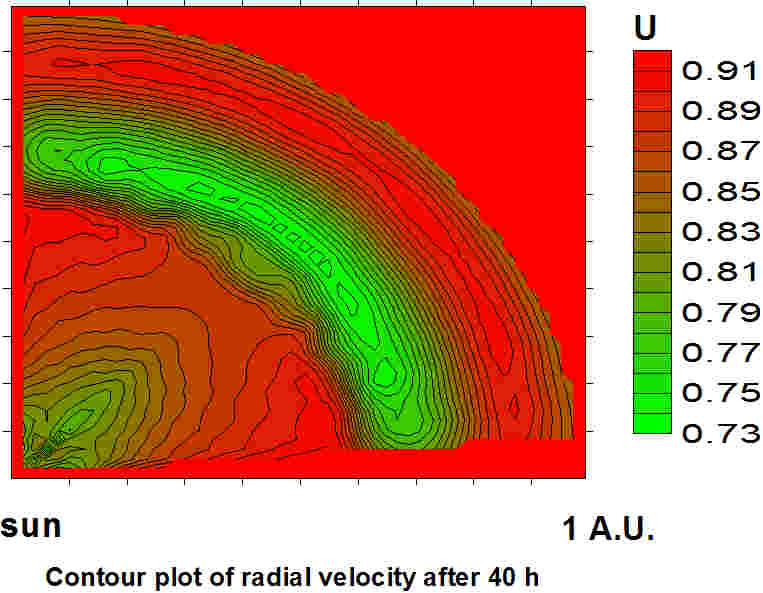}%
\caption{...}%
\label{Fig_unn40Contour}%
\end{center}
\end{figure}

Finally, on Figure \ref{Fig_unn40Contour} we provide the Contour plot of the
\textbf{radial} velocity at $40$ hours after the CME:\ %

\begin{figure}
[!ptb]
\begin{center}
\includegraphics[
height=2.237in,
width=2.8463in
]%
{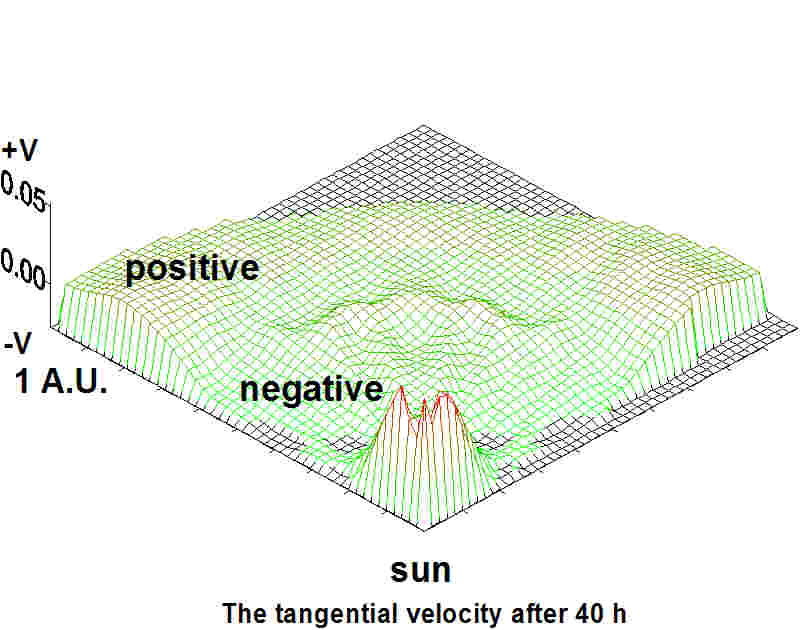}%
\caption{...}%
\label{Fig_vnn40}%
\end{center}
\end{figure}

On Figure \ref{Fig_vnn40} we provide the tangential velocity $V$ of the
disturbed solar wind at $40$ hours after the CME; the values of $V$ are dimensionless:\ %

and on Figure \ref{Fig_40tang} below we provide the contour plot of the
tangential velocity $40$ hours after the CME:%

\begin{figure}
[!ptb]
\begin{center}
\includegraphics[
height=2.1278in,
width=2.7397in
]%
{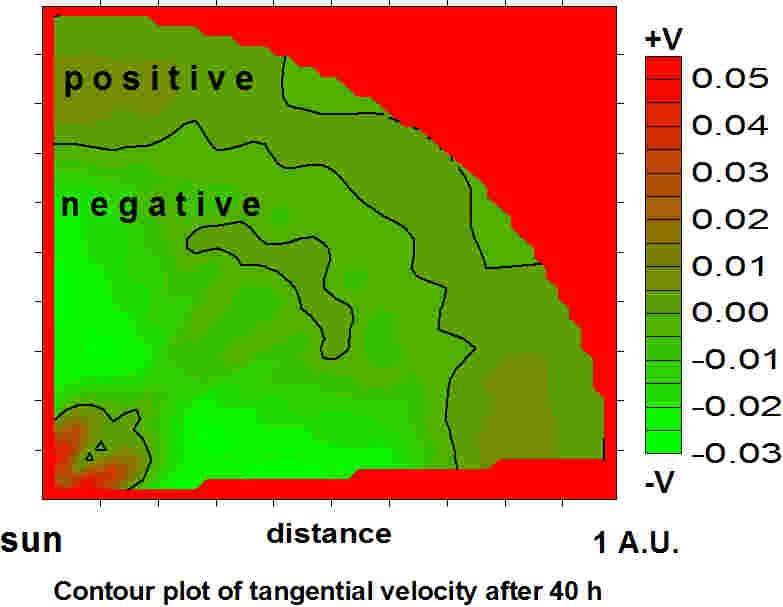}%
\caption{...}%
\label{Fig_40tang}%
\end{center}
\end{figure}

What concerns the magnetic field, we refer to the paper \cite{srebrov2003},
where a detailed figures of the disturbed tangential IMF $B_{z}$ are provided.
In \cite{srebrov2003} it is seen that the tangential component of the magnetic
field vector $B_{t}$ (which in fact coincides with the $B_{z}$ component in
the geocentric Cartesian coordinate system) also has positive and negative
values. Thus a structure is formed in which the direction of the disturbed
magnetic field vector is changing to the north or to the south (in the same
coordinate system mentioned above). As we know these changes are the ones
found to be the main cause of the geomagnetic storms at coupling of the
disturbed interplanetary magnetic field with the Earth's magnetosphere
magnetic field. From the results of the simulations it is visible that the
initial relatively closely disturbed conical area, caused by the CME, also
expands in the tangential direction during the propagation and has values,
which are comparable with the data gathered for example from the observations
near the Earth orbit.

The above explanation justifies why is the component $B_{z}$ of the IMF so
important for the understanding of the Geomagnetic storm.

\subsection{Ground Geomagnetic Field, and geomagnetic activity index during a
storm \label{sectionComponents}}

One of the classical models of geomagnetic storm is the Chapman model,
\cite{akasofu}. It describes the disturbance of the (ground) geomagnetic field
variation during a geomagnetic storm.

According to Chapman's analysis, \cite{akasofu}, if the time $t=0$ denotes the
sudden start of the storm, then at a certain time point $t$ the disturbed
magnetic field vector $D$ measured at a point on Earth's surface (with
components denoted by $D(H),$ $D(D)$ and $D(Z)$ ) can be expressed in a
Fourier series expansion as follows:%
\[
D\left(  \theta,\varphi,t\right)  =c_{0}\left(  \theta,t\right)  +%
{\displaystyle\sum_{n=-\infty}^{\infty}}
c_{0}\left(  \theta,t\right)  \sin\left(  n\varphi+\alpha_{n}\left(
\theta,t\right)  \right)  ;
\]
here $\theta$ is the complement of the geomagnetic latitude to $90^{0}$ ,
$\varphi$ is the geomagnetic longitude, and $\alpha_{n}$ is the phase angle.
The first and the second terms in this expression represent the axially
symmetrical component of the dipole axis and the asymmetric part of the
disturbed field that varies with the longitude $\varphi.$ These components are
referred to respectively \textbf{storm-time variation} ($D_{st}$) and
\textbf{local time-dependent disturbance} ($DS$), which contain daily regular
variation $S_{r}$ of type $S_{q}$ and the variance from the asymmetric part of
the Ring current. So we can write:%
\begin{equation}
D=D_{st}+DS; \label{D=Dst-Defined}%
\end{equation}
here $DS$ is a \textbf{geomagnetic index} that characterizes local storms. The
variation of the horizontal component $D_{st}\left(  H\right)  $ of the
$D_{st}$ field is a function of the variable $\theta.$ It is larger at low
latitudes. The declination $D_{st}\left(  D\right)  $ has little change during
the geomagnetic storm. The vertical component $D_{st}\left(  Z\right)  $ also
changes slightly compared to the horizontal component. Thus, the $D_{st}$
variation is practically parallel to the Earth's surface except in the areas
of the polar cap, where the vertical component has larger positive changes.
For that reason the main interest represents the horizontal component
$D_{st}\left(  H\right)  .$ Hence, we have the formula for the
\textbf{horizontal components}
\begin{equation}
DS(H)=D\left(  H\right)  -D_{st}(H) \label{DS}%
\end{equation}
where usually $D\left(  H\right)  $ is denoted simply by $H\left(  t\right)
,$ and called \textbf{the horizontal component} of the field.

By neglecting $D_{st}\left(  D\right)  $ and $D_{st}\left(  Z\right)  $ as
\textbf{small quantities}, it is apparent from the above analysis that the
magnitude of the geomagnetic storm is determined mainly by the horizontal
component $D_{st}\left(  H\right)  $, and this is the geomagnetic activity
index during the storm. It describes the intensity of the symmetrical part of
the circular current that occurs in the equatorial area of the magnetosphere
during a magnetospheric storm. This is the global part of the geomagnetic
storm index. It represents the mean value of the disturbed horizontal
component of the geomagnetic field determined by data from several low-level
observatories, distributed by geographic lengths. On quiet days, this
variation may be around $\pm20$ \emph{nT}, but during a geomagnetic storm it
reaches large negative values of the order of hundreds of \emph{nT}.

\begin{remark}
The determination of the $D_{st}$ \textbf{index} is provided every hour, and
is practically determined by taking an average of the data through a
consortium of geomagnetic observatories. The process of practical
determination of this geomagnetic index is described in detail in the IAGA bulletin.
\end{remark}

\begin{remark}
Geomagnetic field variation data in the past decades contained mean
\textbf{hour} values, however, the present data contain mean \textbf{minute}
values, and in all INTERMAGNET observatories, the registration is done with
\textbf{mean-second} values. This shows an increase in the information about
the geomagnetic field and one may speak about "big data"-shift of the
measurement paradigm.
\end{remark}

\paragraph{The $D_{st}$ index during the 2003 storm \label{SectionDstGraph}}

\begin{figure}
[!ptb]
\begin{center}
\includegraphics[
height=2.5625in,
width=3.4069in
]%
{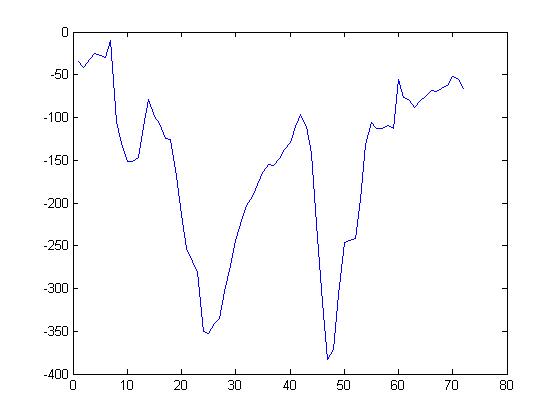}%
\caption{The $D_{st}$ index on October $29$ - November $1,$ $2003.$ }%
\label{FIG_dst1_kyoto}%
\end{center}
\end{figure}

In order to give an idea about the general form of the $D_{st}$ index during
the storm we provide the $D_{st}$ data during the storm in October $29,$
$2003$, downloaded from the WDC (World Data Center) for Geomagnetism in Kyoto;
the data available are for $3$ days, every hour, see Figure
\ref{FIG_dst1_kyoto}. On Figure \ref{FIG_dst1_kyoto} we see that the $D_{st}$
variation has two big decreases due to two different CMEs.%

\subsection{Ionospheric parameters from ionospheric sounding stations}

Let us remind that the ionosphere of the Earth can be seen as a
\textbf{conductive layer}. Its motion induces an electromotive force $v\times
B$. Here $v$ is the vector of the \textbf{drift velocity} of the charged
particles (not to be mixed with the tangential macro-velocity of the solar
wind!), and $B$ is the geomagnetic field vector in the \textbf{ionosphere}.

Let us mention that the ionosphere is influenced by different factors which
cause some \textbf{long-period (low frequency)} ionospheric parameter changes:

\begin{enumerate}
\item \textbf{Solar} \textbf{tidal} movements and movements, caused by the
periodic warming of the atmosphere by the Sun, cause changes in the
geomagnetic field, called quiet-solar variations. The latter, as noted above,
are referred to as $S_{q}$. In turbulent conditions, we have a regular
variation $S_{r}$ associated with the local time, which differs from $S_{q}$
due to the influence on the magnetospheric storm ionosphere. As with $S_{q}$,
the conditions in the high atmospheric layers are particularly important and
essential. For example, the concentration of the charged particles in the
ionosphere, which depends on UV ionization, as well as the degree of invasion
of charged particles in the ionosphere from the above areas, such as the
plasosphere and the magnetosphere.

\item The distribution in the ionosphere of the electromotive forces,
determined by the \textbf{lunar} \textbf{tidal} movements of the atmosphere,
is approximately fixed in relation to the moon. Thus, the induced current and
the resulting magnetic field is also fixed for an observer on the moon. Each
magnetic observatory performs one turn per day about this distribution,
turning in a circle defined by the latitude. Therefore, the stations register
a variation of the geomagnetic field over time. This variation is called
lunar-day and is denoted by $L$.
\end{enumerate}

\subsubsection{Data about the parameters of the ionospheric plasma}

In the present work we have used data for different parameters of
\textbf{ionospheric plasma} in a specific local area, as TEC, F2, etc. In the
experiments presented we have limited ourselves with the TEC data since it is
considered as the most important of all parameters. For example, we use the
ionosphere sounding data from ground ionospheric stations, which are chosen to
be located near the (ground) geomagnetic field registration points. This
allows by comparison of the two "signals" to seek for the presence or absence
of a possible correlation between them. Thus, we have the possibility of
identifying the origin of individual modes (wave packages) and groups of modes.

The \textbf{ionospheric} and geomagnetic data are synchronized in universal
time so that we can monitor for the presence or absence of simultaneity of the
two signals. We have used data on ionospheric parameters with a sampling
frequency comparable to the geomagnetic data, namely, every $5$ minutes. This
is very interesting in terms of decomposition of geomagnetic variations. Of
course, the \textbf{modes} associated with extra-ionospheric origin can also
be identified and this has already been commented (as for example the
geomagnetic pulsations).

Let us note that in recent years, the \textbf{ionospheric} data registration
also tends to increase the sampling frequency and the sampling has now reached
in some stations a period of $5$ minutes between two samples. Given that a
large number of ionospheric plasma parameters are measured at these stations,
we can assume that they may also be referred to as "Big Data" paradigm.

\begin{figure}
[!ptb]
\begin{center}
\includegraphics[
height=2.2524in,
width=3.634in
]%
{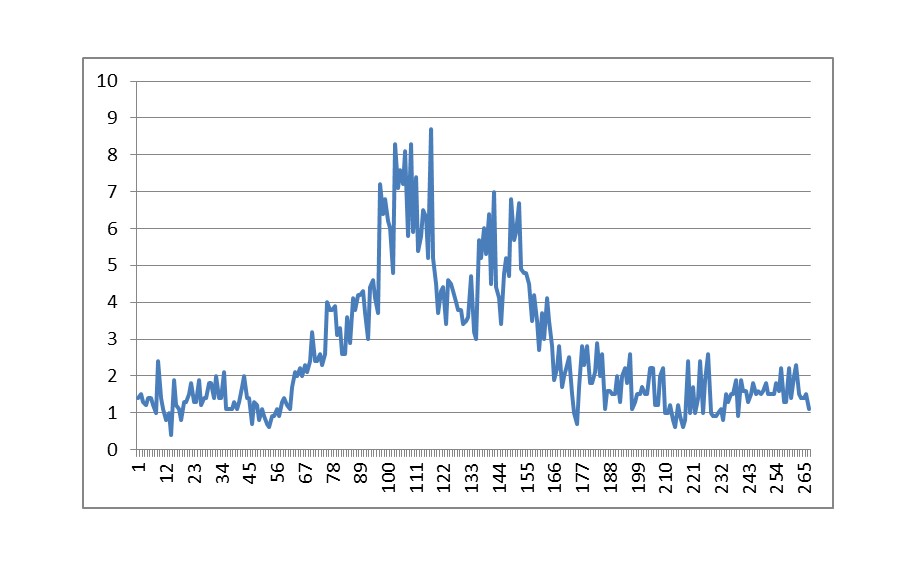}%
\caption{Graph of the TEC data from the ionosound station in Athens, on
January $4,$ $2018.$}%
\label{FIG_TEC_January4_2018}%
\end{center}
\end{figure}

For example, on Figure \ref{FIG_TEC_January4_2018} below we provide a graph of
the TEC data from the ionosound station in Athens, on a quiet day, January 4,
2018. The variation of the TEC data is shown. We see that the main trend is
given by the daily variation of the TEC. The fast oscillations with periods
less than $3$ hours are observed during the whole day. At midday time we
observe modes (wave packages) with a larger amplitude which are apparently of
soliton type. This phenomenon has been studied in \cite{Belashov2015}.%

\begin{remark}
In an interesting research of the short period modes (wave packages) in the
Ionosphere, V. Belashov has created soliton model for their explanation, see
\cite{Belashov2015} and references therein; matching of the model to the TEC
values is presented on p. $338.$ See also \cite{SrebrovPashovaKounchev}.
\end{remark}

\subsection{Emergence of higher-frequency modes in the Ionospheric parameters
and in the IMF which are related to the ground Geomagnetic field variations}

Up to here we have considered the long-periodic Geomagnetic variation
phenomena were considered above. Below we describe shortly the generation of
short-period (high-frequency) Geomagnetic variations caused by
\textbf{low-frequency plasma instabilities}.

As already mentioned above, the plasma in the ionosphere, plasmospher and the
interplanetary colisionless plasma medium have considerable instability.
Instability is also occurring in the radiation belts of the Earth, i.e. in the
Magnetosphere although most of the latter do not realize the conditions
characterizing the medium as plasma. Some examples of cosmic plasma show the
presence of instabilities that produce non-thermal waves and various
distribution functions of kinetic particles parameters, especially in the
ionosphere and the plasmosphere.

\textbf{Instabilities} can also cause a non-linear effect of wave propagation
in the ionosphere, plasmosphere and interplanetary medium, and they also cause
collisions and acceleration of fast particles in astrophysical plasma.

For the current work, it is important to note the low-frequency instabilities
in terms of natural plasma frequencies (MHD instabilities, fluid instabilities
and drift instabilities). They create the low-frequency modes in the
Geomagnetic field and in the Interplanetary magnetic field. The properities of
the wave modes are strong functions of frequency. For low-frequency plasma
modes, the circular frequency of the waves is much smaller than the natural
frequencies as the plasma frequency and the cyclotron frequency of the plasma.

Practically, very often these \textbf{low-frequency} plasma modes coincide
with the considered by us \textbf{higher-frequency} geomagnetic field
variations. In this work, we use geomagnetic field registration data on the
ground with periods more than $1$ minute, that are associated with
\textbf{low-frequency modes} in the Earth's plasma cosmic environment, and
also with Ring current fluctuations in the Magnetosphere.

In \textbf{plasma}, the macroinstabilities occur in the low-frequency mode and
usually involve the magnetic field (not just the ground geomagnetic field!).
Therefore, in the \emph{present research} we use data about the variations of
the\textbf{ }magnetic field in different areas of the near Earth space and
those generated in the interplanetary space.

What concerns the \textbf{short period (higher-frequency)} ionospheric
parameter variations, let us recall that the variation can be decomposed, as a
superposition of sources located in the magnetosphere and the ionosphere and
also associated with the various tidal movements of the high Atmosphere. The
high frequency (ground) geomagnetic field variations may be a result of short
period Ionospheric parameters variations.

It is clear from the above that it is \textbf{necessary }to use data on the
\textbf{ionosphere} status as well as data of the \textbf{interplanetary
magnetic field (IMF), }and the ground Geomagnetic field during the storm.

In the present work, in each case, we choose to study the behavior of the
various magnetic field components that are associated with certain processes
in the respective area of observation.

In the Interplanetary medium disturbances of the \textbf{IMF} and of the solar
wind propagate, as we have already mentioned above. Beyond that, in the
Interplanetary medium different \textbf{low-frequency} plasma modes are
generated, as a result of the plasma instabilities. Data on \textbf{IMF}
variations induced by macro-instability of the interplanetary plasma medium
are recorded by satellites located outside the Earth's magnetosphere, as for
example the \textbf{satellite ACE} whose data we use. These data contain the
values of the components of the magnetic vector as well as the values of its
magnitude, measured by different instruments. In order to study these
instabilities and related waves in the interplanetary medium, we use data for
the IMF with data sampling $4$ minutes. Magnetic modes in these media and in
this frequency range are associated with the so-called macroinstability.

Finally, \textbf{one of the main objectives of our study} is to identify modes
with \textbf{short periods }(with frequencies much larger than the natural
plasma frequencies), which are generated by microinstabilities, and are caught
by the ground geomagnetic field registration and in the plasma ionospheric parameters.

\subsection{The strong geomagnetic storms in $2003$ and $2017$ to be analyzed
\label{sectionDisturbances}}

In the present research we apply Wavelet Analysis to analyze short-period
Geomagnetic Field variations, Ionospheric parameter variations, and
Interplanetary magentic field variations, during the manifestation of the two
famous strong geomagnetic storms during the $23$rd solar cycle in $2003,$ and
during the $24$th solar cycle in $2017.$

\subsubsection{The storm in 2003}

The data obtained from Solar Astronomy observations have provided the
following report:

During the $23$rd solar cycle in October and November $2003$ there were two
very strong storms. One started on October 29 and the other began on November
$21$. In the last ten days of October 2003, the lean activity has gone to an
extremely high level. On October 18, a large active region (AR), turning north
of the solar equator, was designated by NOAA as AR 484. On October 28, AR 484
was located near the sub-Earth point of the solar disk $8^{0}$ on the east of
the central meridian and $16^{0}$ north latitude. At 11:10 UTC AR 484 produced
one of the largest solar flares for the current solar cycle. This flare was
classified as X17 (peak X-ray flux $1.7\times10^{-3}$ W/m$^{2}$ ).

An extreme CME with a radial plasma velocity of $2500$ km/s was observed. The
mass ejected from this CME was in the range of $1,4-2,1\times10^{13}$ kg, and
the kinetic energy released was $4,2-6,4\times10^{25}$ J. The following day,
October 29, AR 484 again produced a large eruption. This peak was X10 (X-ray
flux $10^{-3}$ W/m$^{2}$ ) at $20:49$ UTC. I was targeting the Earth halo at a
speed of $2000$ km / s and kinetic energy of $5,7\times10^{25}$ J. The
interplanetary magnetic field (IMF) reached about -50 nT, its normal value, in
calm conditions, is ten times lower. The shock wave of the event on Oct. 28
was determined by the Advanced Composition Explorer (ACE) spacecraft at
$05:59$ UTC. At $06:13$ UT was registered an SSC pulse, marking the beginning
of the sixth storm by the registration stamp (since 1932). On 29 and 30
October the planetary index $Kp$ reached value 9. The geomagnetic storm
continued until November 1 and had a horizontal component down to around -400
nT. The highest value of the $D_{st}$ index was registered on October 30 at
$23:00$ UT.

\subsubsection{The storm on 7-8 September, 2017}

The other storm considered in the present work is the one on September $7-8,$
$2017$. It was one of the most flare-productive periods of now-waning solar
cycle $24$. Solar active regions (AR) $2673$ and $2674$ both matured to
complex magnetic configurations as they transited the disk. AR2673 transformed
from a simple sunspot on 2 September to a complex region with
order-of-magnitude growth on 4 September, rapidly reaching beta-gamma-delta
configuration. In subsequent days the region issued three X-class flares and
multiple partial halo ejecta. Combined, the two active regions produced more
than a dozen M-class flares. As a parting shot AR2673 produced: 1) an X-9
level flare; 2) an associated moderate solar energetic particle event ;and 3)
a ground level event, as it arrived at the solar west limb on 10 September.
From $4-16$ September the radiation environment at geosynchronous orbit was at
minor storm level and $100$ MeV protons were episodically present in
geostationary orbit during that time frame. The early arrival of the coronal
mass ejection associated with the 6 September X-9 flare produced severe
geomagnetic storming on 7 and 8 September. The full set of events was
bracketed by high speed streams that produced their own minor-to-moderate
geomagnetic storming.

\subsection{Acquired Data for short period variations of the Geomagnetic
field, the Ionospheric parameters, and the IMF,\label{sectionDataSources}}

As we have explained in section \ref{SectionMechanism}, the global picture of
the geomagnetic phenomena is very complicated and dynamic. For that reason,
for the explanation as well as for the prediction of its dynamics one needs to
attract as much as possible observable data, which form the basis of our Big
Data analysis.

We analyze high frequency time series data from different sources. Let us
remind that "high frequency" registrations in Geomagnetism are of the order
$0.1-10$ mHz (i.e. of periods $1,66$ min till $2,77$ hours). The following
three high frequency time series were acquired:

\begin{enumerate}
\item $T_{G}:$ Time series for the ground Geomagnetic data (from ground
geomagnetic observatories, $1$ min. sampling and $1$ sec. sampling), in nT.
Our main objective is to seek for correlations in the wavelet coefficients of
the CWT of the above time series which explain the dynamics of different
geomagnetic phenomena.

\item $T_{IP}:$ Time series for the Ionospheric parameters - TEC (from
ionospheric sounding stations, $5$ min. sampling), in TEC unit.

\item $T_{IMF}:$ Time series for the Interplanetary Magnetic Field (from the
ACE satellite, $4$ sec. sampling), in nT.
\end{enumerate}

\subsection{Data about the strongly disturbed geomagnetic field in October and
November 2003 and September 2017}

\subsubsection{The H-component of the geomagnetic data from
\textbf{Panagiurishte (}PAG) observatory}

\begin{figure}
[!ptb]
\begin{center}
\includegraphics[
height=2.59in,
width=3.4486in
]%
{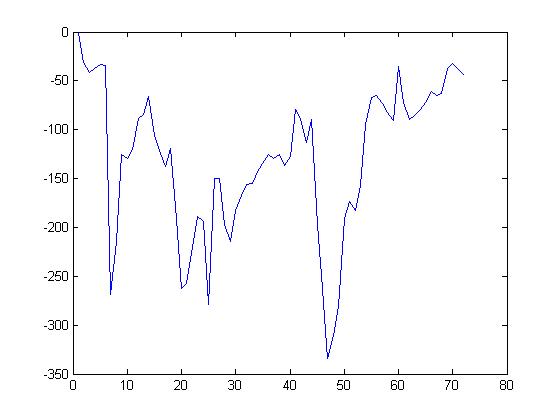}%
\caption{The mean-hour values of the $H-$component registered at PAG, October
$29-31,$ $2003.$}%
\label{FIG_v1_panag_storm1}%
\end{center}
\end{figure}

On Figure \ref{FIG_v1_panag_storm1} below we have the variation of the
$H-$component (see formula (\ref{DS})) registered at the geomagnetic
observatory PAG during the geomagnetic storm on October $29,$ $2003.$ The data
are mean-hour values, registered from $0$ h on October $29$ till $24$ h on
October $31.$%

\subsubsection{The $DS$ index from the Surlary (SUA) geomagnetic data}

On Figure \ref{FIG_ds_sua_min_storm1__} below, we provide the data for the
index $DS$ variation (defined in formula (\ref{D=Dst-Defined}) during the
storm $29-31$ October, $2003,$ registered at the Surlary (SUA) geomagnetic
observatory in Romania. The sampling of the data is $1$ minute.%

\begin{figure}
[!ptb]
\begin{center}
\includegraphics[
height=2.5585in,
width=3.4395in
]%
{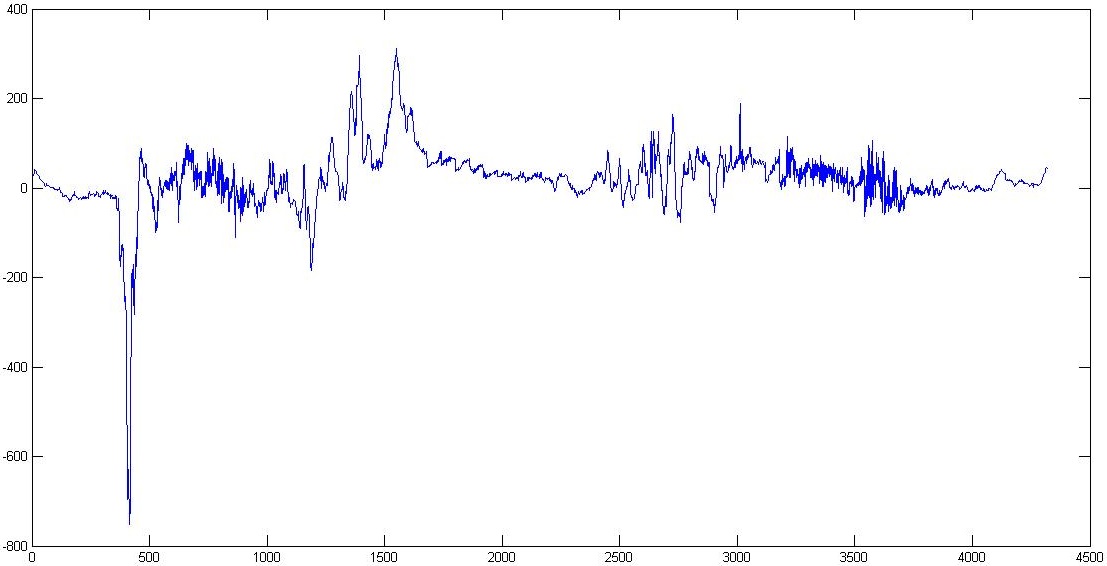}%
\caption{ The index $DS$ variation during the period $29-31$ October, $2003,$
registered at the Surlary (SUA). The sampling is $1$ minute. }%
\label{FIG_ds_sua_min_storm1__}%
\end{center}
\end{figure}
We see all details due to the fact that the data are provided every single minute.

\begin{remark}
It is questionable whether it is worth applying the CWT to the $DS$ index
given by $H\left(  t\right)  -D_{st}$ (see formulas (\ref{D=Dst-Defined}) and
(\ref{DS})), or directly to the rough data of the $H-$component, since we are
in principle seeking for variations of $H,$ but $D_{st}$ index is provided on
hourly basis. This might create artificial jump every hour.
\end{remark}

\section{EXPERIMENTS \label{SectionExperiments}}

In the experiments to follow, we are motivated by the interest to discover
wave packages of short periods and their correlations in the three different
sources of data which we have already discussed:

\begin{enumerate}
\item (ground) Geomagnetic data, from geomagnetic observatories

\item Ionospheric data (TEC values, from ionosound stations)

\item IMF data (from satellite ACE),
\end{enumerate}

We would like to identify any kind of correlation and causality among them by
applying the method of Wavelet Analysis.

\subsection{References on applications of Wavelet Analysis to
Geomagnetism\label{sectionWaveletGeomagnetism}}

Before presenting the results of our experiments, we provide some references
which might be useful to the reader.

Let us note that in a number of works, \cite{Mandrikova}, \cite{Wei},
\cite{artigas}, \cite{boudouris}, \cite{Jach}, \cite{katsavrias}, \cite{xu},
the variation of geomagnetic data (in particular of $D_{st}$) is analyzed by
means of wavelet analysis. Recently, wavelet analysis of geomagnetic field
perturbations was widely used in the study of tsunami waves
\cite{Klausner2011}, \cite{Klausner2014}, \cite{Klausner2016},
\cite{Klausner2016b}, \cite{Klausner2017}, \cite{Schnepf2016}. In \cite{Jach},
\cite{xu2011} wavelet analysis of the geomagnetic field is used to define a
new index, alternative to $D_{st},$ but on a minute basis.

\subsection{Experiments with data on a quiet day, 28 July, 2018}

In order to have controls over the statistical behavior of the geomagnetic
data during geomagnetic storm, we have taken data for quiet days from two
geomagnetic observatories -- in situ repeat station at Balchik (Bulgaria), and
from another geomagnetic observatory Surlary (SUA), in Romania; the distance
between them is about $190$ km. This means that they have almost identical
conditions, and it is well known that there are no strong magnetic variations
from natural or artificial character in the regions.

\subsubsection{Visualization of the Wavelet Analysis}

We have provided a brief summary on the Continuous Wavelet Transform in
Appendix, section \ref{AppendixWavelets}.

In the experiments below we perform CWT to time series $f\left(  j\right)  ,$
where $f\left(  j\right)  =F\left(  \delta t_{j}\right)  $ for some
"continuous time series" $F\left(  t\right)  ,$ and $t_{j}\delta$ are the
sampling times on a uniform mesh, $t_{1},t_{2},...,t_{N}$. Here $\delta$
denotes the sampling interval, for example, we have $\delta=1$ second, $4$
seconds, $1$ minute, $1$ hour, etc. We visualize the absolute values
$\left\vert W_{\psi}f\left(  a,b\right)  \right\vert $ of the CWT coefficients
$W_{\psi}f\left(  a,b\right)  $ defined in formula (\ref{CWTdefinition}). In
all our experiments the shifts $b$ (visualized on the $X-$axis) run through
the full set of indexes $1,2,...,N.$ On the other hand, the non-negative
parameter $a$ in formula (\ref{CWTdefinition}) which denotes the scaling
parameter (and has the meaning of \textbf{periodicity}), is interesting for us
mainly for shorter intervals. Hence, in some experiments we consider only
periods $a\leq N_{1}$ for some maximal period $N_{1}<N.$ The parameter $a$ is
visualized on the $Y-$axis. In order to get a better idea of the behaviour of
the CWT coefficients $W_{\psi}f\left(  a,b\right)  ,$ we find it instructive
to have the visualization of $\left\vert W_{\psi}f\left(  a,b\right)
\right\vert $ both as a \textbf{heatmap} and as a \textbf{contour} map, see
for example, Figure \ref{FIG_balchik_28_7_2018_quiet_5sec} below.

\subsubsection{Experiments with Balchik Geomagnetic data, $28$ July, $2018,$ 1
second data}

The $H-$component geomagnetic data were collected every second, for 24 hours,
at a repeat station in Balchik (Bulgaria), on $28$ July, $2018.$

We provide the graph of the time series of the data Figure
\ref{FIG_balchik_h-component_2018_28_july_1sec} below. The daily variation of
the field is clearly visible as the main trend, and also some rather permanent
short-period variations.%

\begin{figure}
[!ptb]
\begin{center}
\includegraphics[
height=2.5569in,
width=3.8566in
]%
{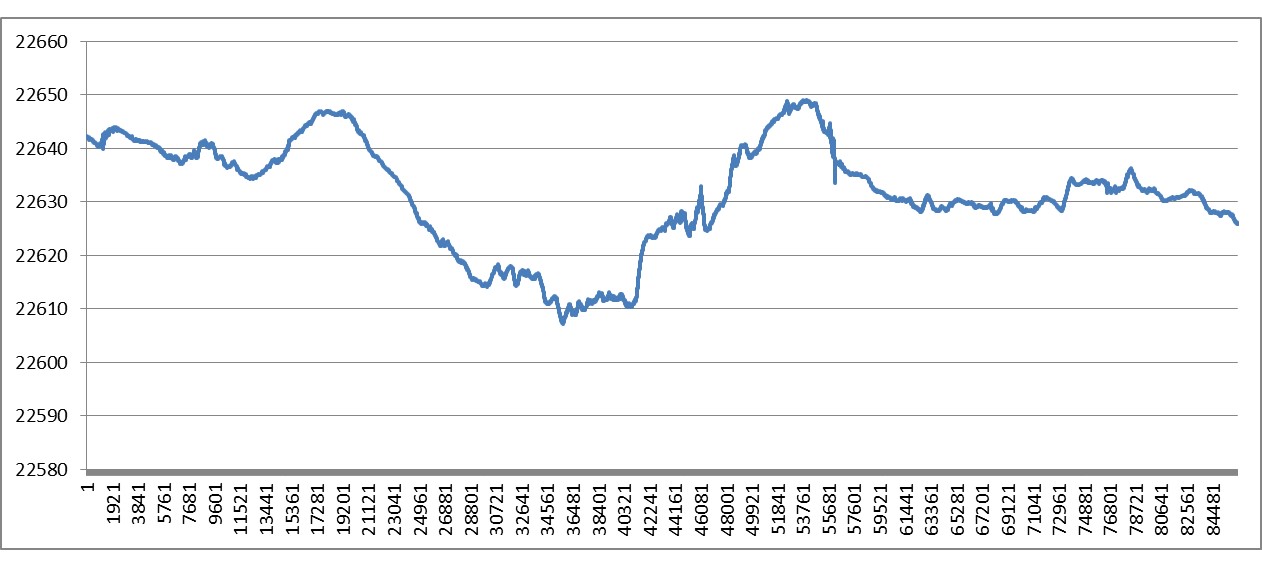}%
\caption{The $H-$component with sampling $1$ second, at a repeat station in
Balchik on $28$ July, $2018.$}%
\label{FIG_balchik_h-component_2018_28_july_1sec}%
\end{center}
\end{figure}

Below, on Figure \ref{FIG_balchik_28_7_2018_quiet_5sec}, we provide the
experiments, namely, the \textbf{heatmap} (on top)\ and the \textbf{contour}
map (on bottom) of the absolute values of the CWT coefficients $\left\vert
W_{\psi}f\left(  a,b\right)  \right\vert $ of the $H-$component .%

\begin{figure}
[!ptb]
\begin{center}
\includegraphics[
height=3.0in,
width=5.0in
]%
{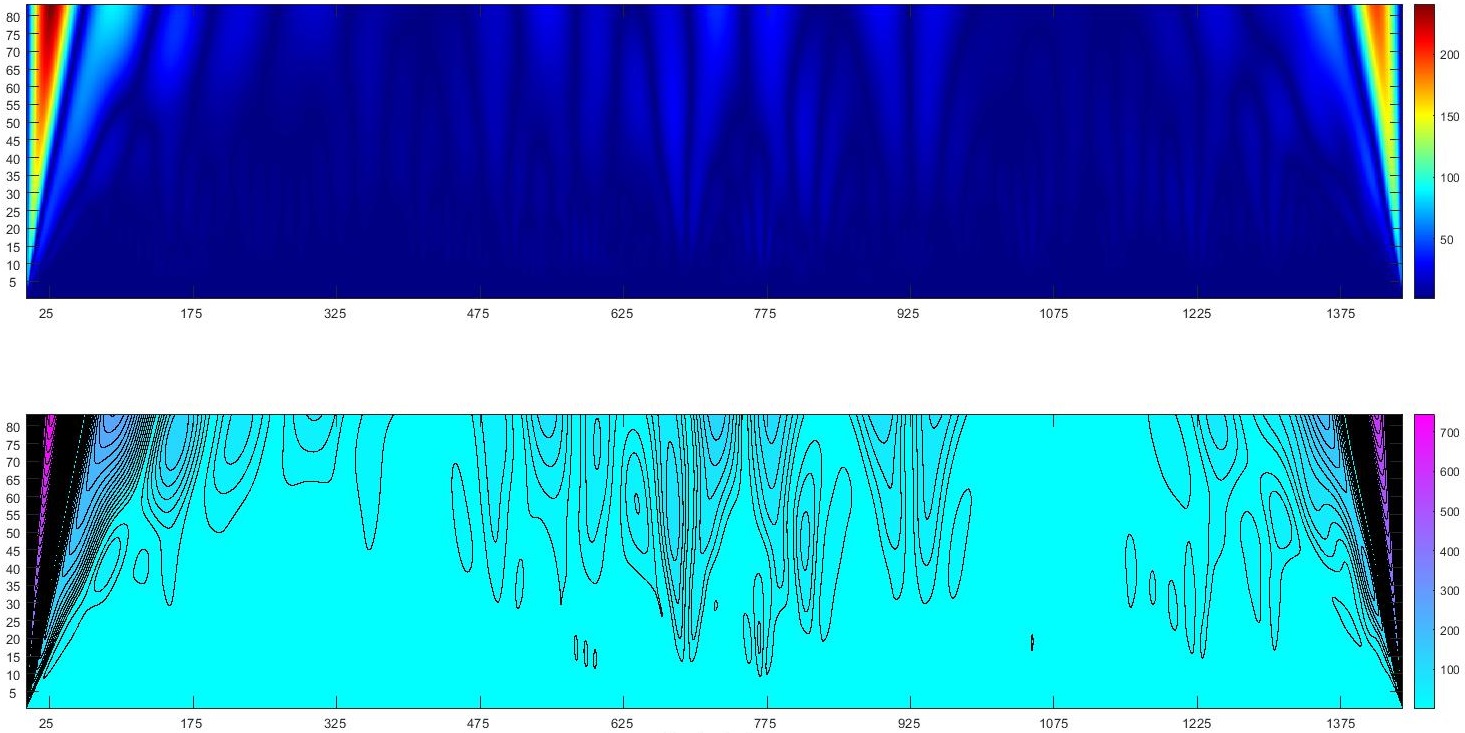}%
\caption{\textbf{Heatmap} (on top)\ and the \textbf{contour} map (on bottom)
of the CWT of the $H-$component of the Balchik data, on $28$ July, $2018$. }%
\label{FIG_balchik_28_7_2018_quiet_5sec}%
\end{center}
\end{figure}

On Figure \ref{FIG_balchik_28_7_2018_quiet_5sec} we see that in the CWT of the
time series of the Balchik geomagnetic data, some very interesting details are
identified. Around midday, some wave packages with periods about $100$ min are
clearly visible. They show the possibility for soliton like oscillations ,
related to the solar terminator, theoretically studied in \cite{Belashov2015},
and which has been recently observed in the Wavelet Analysis experiments with
geomagnetic data in the paper \cite{SrebrovPashovaKounchev}.

\subsubsection{Experiment with SUA geomagnetic data on 28 July, $2018$}

\begin{figure}
[!ptb]
\begin{center}
\includegraphics[
height=3.0in,
width=5.0in
]%
{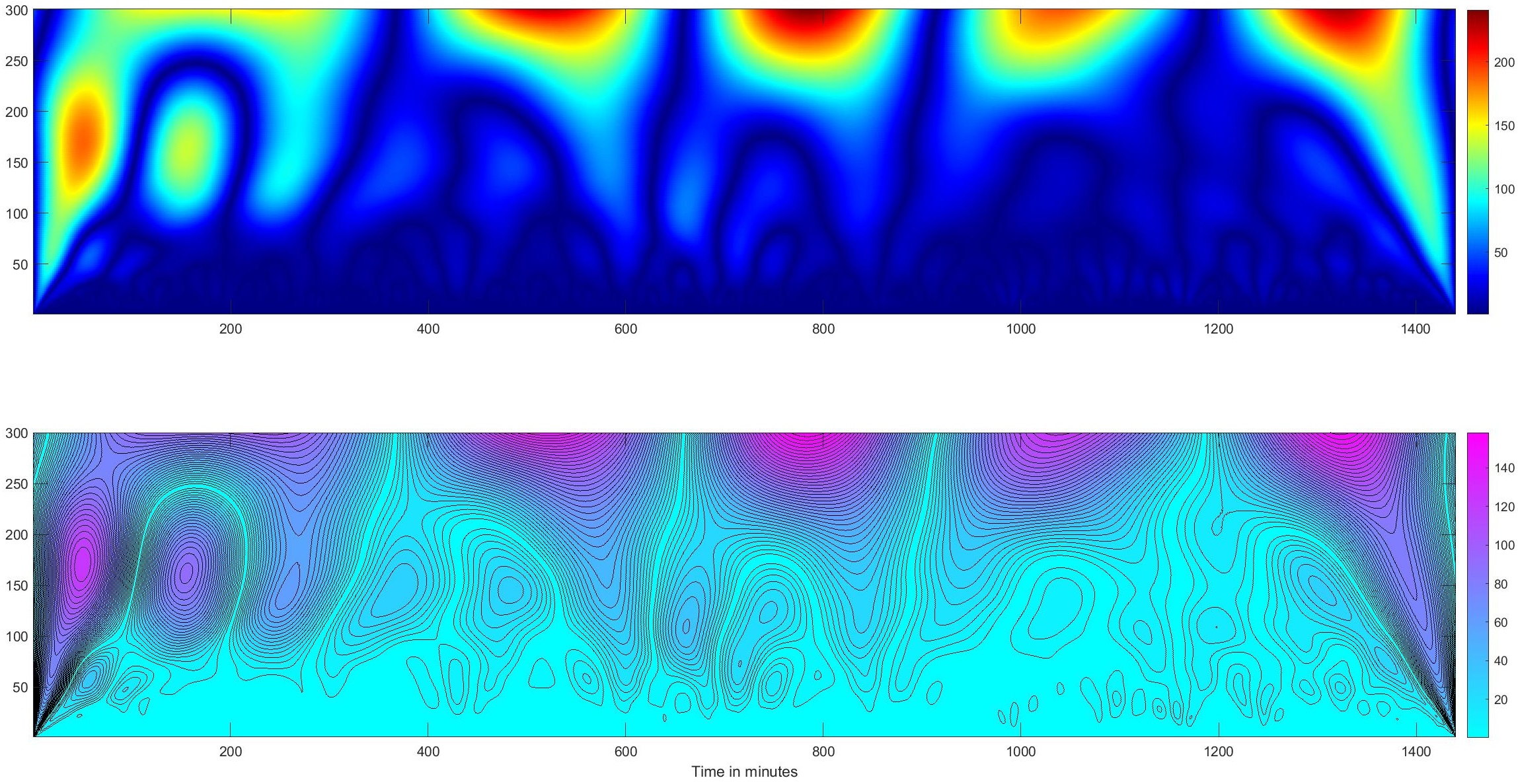}%
\caption{\textbf{Heatmap} (on the top) and the \textbf{contour} \textbf{map}
(on the bottom) of the CWT of SUA data, July $28,$ $2018.$ }%
\label{FIG_sua_28_7_2018_quiet_hcomp_1min}%
\end{center}
\end{figure}

We have made experiments with the time series formed by the $H-$component of
the geomagnetic data from SUA (Surlary, Romania), on whole day, July 28,
$2018$; the sampling is every minute. The \textbf{heatmap} (on the top) and
the \textbf{contour} \textbf{map} (on the bottom) of the CWT is provided on
Figure \ref{FIG_sua_28_7_2018_quiet_hcomp_1min}.

%

\begin{remark}
On the Figures \ref{FIG_sua_28_7_2018_quiet_hcomp_1min} we see wave packages
of $100$ minutes during the whole day, and wave packages with periods below
$180$ minutes about midday. At midday one observes an intensive process of
generation with a period between $100$ to $150$ minutes. Right at the same
time interval (between $600$ and $700$ minutes) there are wave packages with
period $30-50$ minutes. One may suggest that the latter phenomenon may be
generated by the solar terminator, and have a soliton structure,
\cite{Belashov2015}.
\end{remark}

\begin{remark}
There is a lot of similarity at the scale of 20-60 minutes between the CWT of
Balchik data and the CWT of SUA data:

We compare the CWT of the Balchik data in Figure
\ref{FIG_balchik_28_7_2018_quiet_5sec}, and the CWT of the SUA data in Figure
\ref{FIG_sua_28_7_2018_quiet_hcomp_1min}. We see that the wave packages with
periods about $80$ minutes are much more expressed in the Balchik second data,
than on the SUA minute data. This shows that the oscillation phenomena carry
persistent character. In particular, as was concluded in the paper
\cite{SrebrovPashovaKounchev}, we may suggest the existence of soliton like
patterns at the periods of 40 to 60 min.
\end{remark}

\begin{remark}
\textbf{CONCLUSIONS}: Wave packages with periods from 10 to 100 min. exist
during quiet geomagnetic days, which are however predominant at midday.
\end{remark}

\subsection{Experiments with data for the Geomagnetic storm on $7,$ $8$
September, $2017$}

In the following we provide a Wavelet Analysis of the data from ground
geomagnetic field, Ionospheric parameters, and IMF, collected during this
strong geomagnetic storm.

First we provide the picture of the main trend which is determined by the
$D_{st}$ index.

\subsubsection{The $D_{st}$ for the period 7-10 Sept., 2017}

This Figure shows the $D_{st}$ graph for the period, Figure
\ref{FIG_dst_7-10sept_2017}:

%

\begin{figure}
[!ptb]
\begin{center}
\includegraphics[
height=2.5in,
width=3.8in
]%
{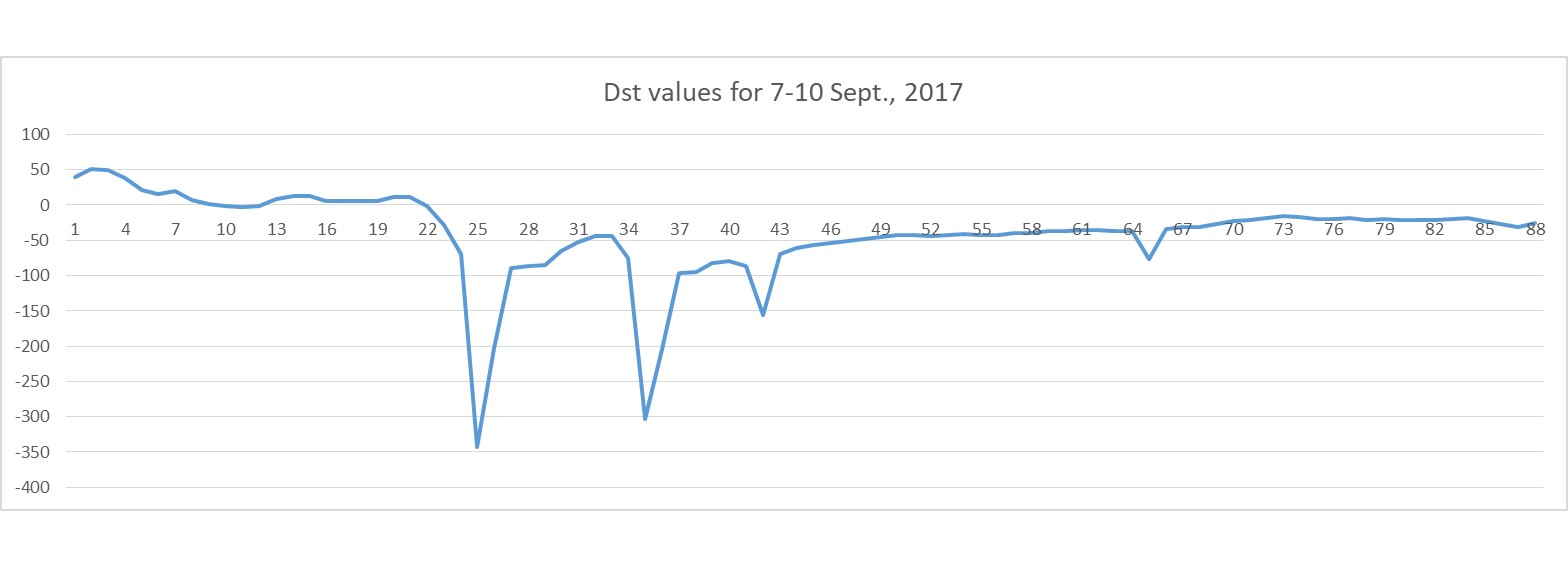}%
\caption{The $D_{st}$ graph for $7-10$ Sept., $2017$. }%
\label{FIG_dst_7-10sept_2017}%
\end{center}
\end{figure}

As we have described the $D_{st}$ index in section \ref{sectionComponents}, it
shows a very unusual behaviour after the storm of 7-10 Sept., 2017.

\begin{remark}
On Figure \ref{FIG_dst_7-10sept_2017} we see the variation of the $D_{st}$
index during the manifestation of the geomagnetic storm in September, $2017.$
On the Figure we see the two decreases of the magnetic field, caused by two
event on the Sun surface, and also a very long recovery phase of the storm
during the period $9-10$ September. This long recovery phase is most probably
related to the lack of short-period variations in the geomagnetic records in
the observatory on September $9$ and $10.$ The decay of the Ring current
cannot create significant geomagnetic variations on the ground in the region
of PAG observatory. However, as we have remarked above, the ionospheric
macroinstabilities during these two days cannot create a variation in the
ionospheric current system, which itself would create variations to be
registered by this ground observatory.
\end{remark}

\subsubsection{Experiments with Geomagnetic data from PAG, 7-10 Sept., 2017, 1
minute data}

\begin{figure}
[!ptb]
\begin{center}
\includegraphics[
height=3.0in,
width=5.0in
]%
{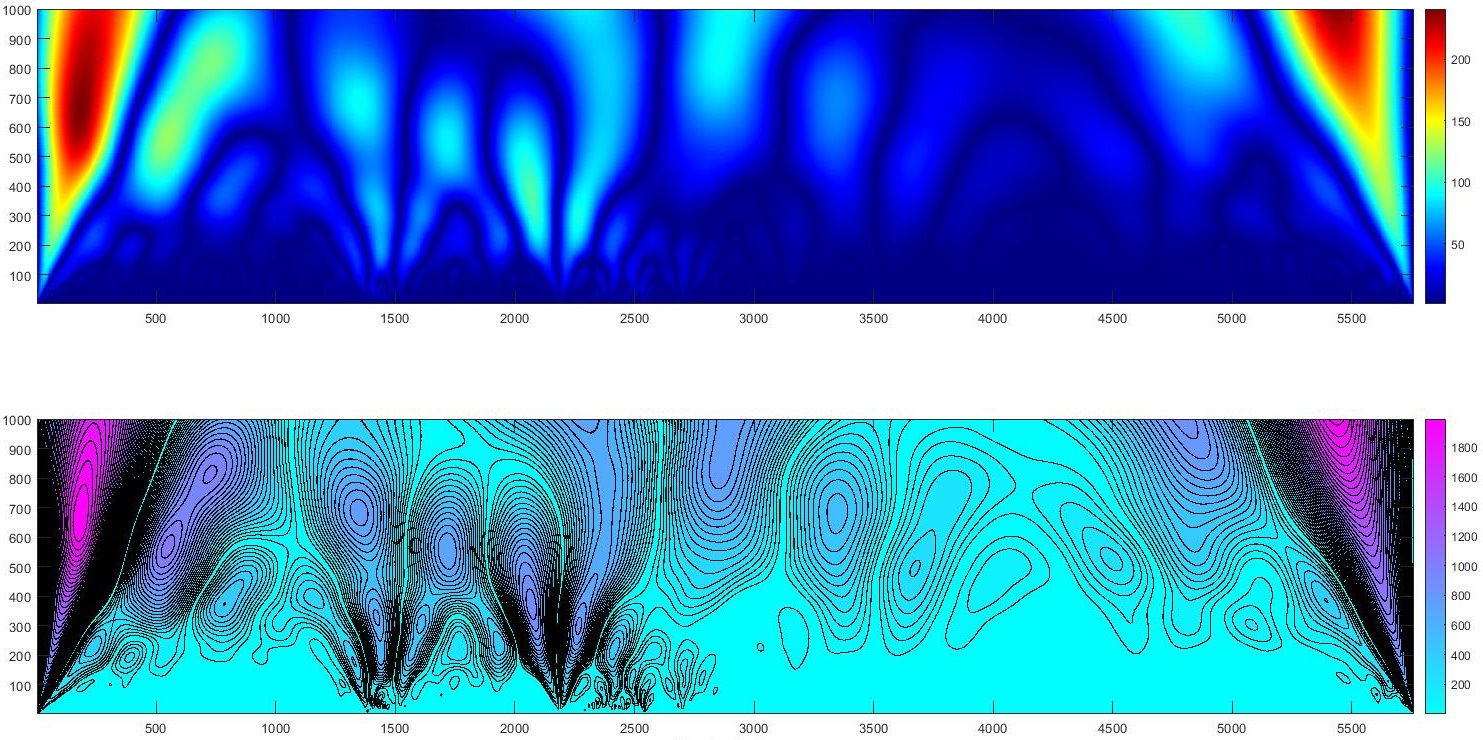}%
\caption{\textbf{Heatmap} (on the top) and the \textbf{contour} \textbf{map}
(on the bottom) of the CWT for $H-$component of PAG, $7-10$ Sept., $2017.$}%
\label{FIG_geomag_panag_1min_2017_7-10_sept_sym8_1000}%
\end{center}
\end{figure}

We have acquired the $H-$component of the geomagnetic data from the
Panagyurishte (PAG) geomagnetic observatory. These data are every 1 minute
sampling period. We provide the \textbf{heatmap} (on the top) and the
\textbf{contour} \textbf{map} (on the bottom) of the CWT, see Figure
\ref{FIG_geomag_panag_1min_2017_7-10_sept_sym8_1000}.

%

We see that during $7$ and $8$ September of the geomagnetic storm we may
identify wave packages with periodicity $20-100$ minutes. However, on $9$ and
$10$ September, one cannot identify wave packages with periods in the interval
$20-100.$

\subsubsection{Experiments with Ionosperic data from Athens, 7-10 Sept., 2017,
$5$ min. data}

We have taken the ionosound TEC data from the ionosound station in Athens
(ATN). The data are measured for full four days 7-10 Sept., 2017, every 5
minutes (this frequency is the modern standard for sampling of ionosounding data).

\begin{figure}
[!ptb]
\begin{center}
\includegraphics[
height=3.0in,
width=5.0in
]%
{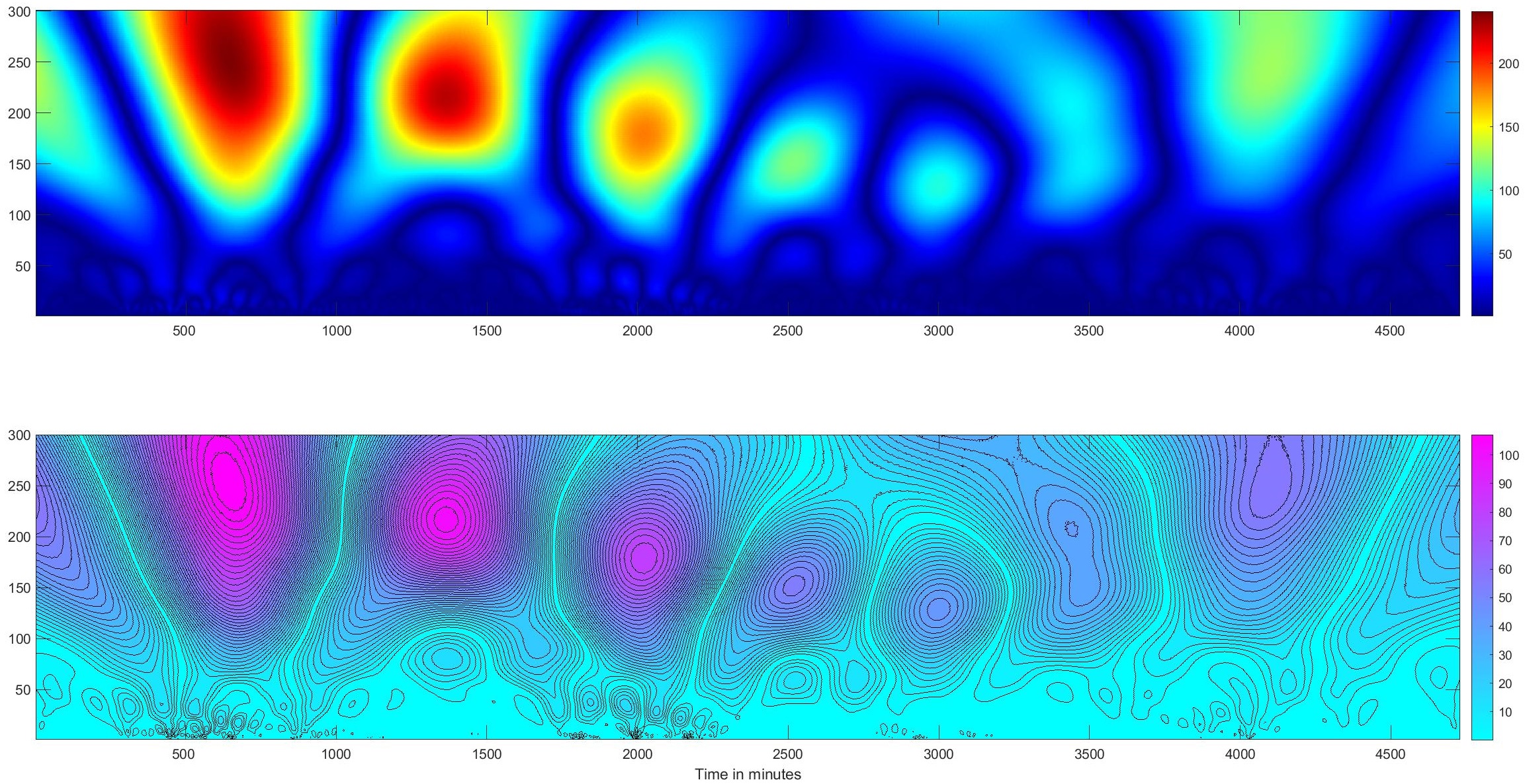}%
\caption{\textbf{Heatmap} (on the top) and the \textbf{contour} \textbf{map}
(on the bottom) of the CWT of the TEC data, Athens, $7-10$ Sept., $2017.$}%
\label{FIG_iono_tec_athens_2017_sept_7-10_storm_5min}%
\end{center}
\end{figure}

On Figure \ref{FIG_iono_tec_athens_2017_sept_7-10_storm_5min} we provide the
\textbf{heatmap} (on the top) and the \textbf{contour} \textbf{map} (on the
bottom) of the CWT of the TEC time series.%

We see that the short period scales which are interesting for us really show
regular patterns. For that reason we have restricted the scales \textbf{only
to 50}, and we show the result below, Figure
\ref{FIG_iono_tec_athens_2017_sept_7-10_storm_5min__50period}:

%

\begin{figure}
[!ptb]
\begin{center}
\includegraphics[
height=3.0in,
width=5.0in
]%
{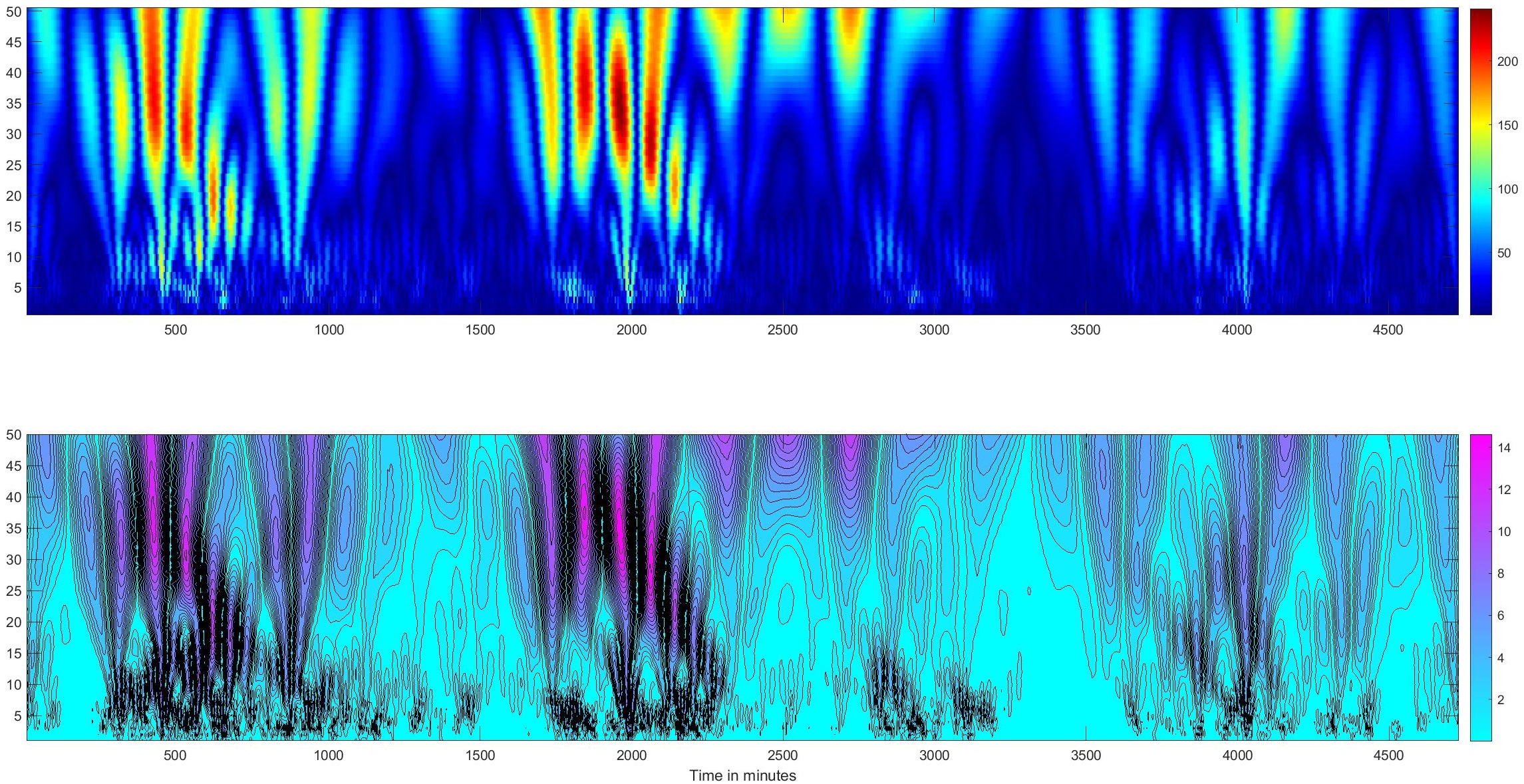}%
\caption{\textbf{Heatmap} (on the top) and the \textbf{contour} \textbf{map}
(on the bottom) of the CWT of the TEC data, Athens, $7-10$ Sept., $2017,$
limited to short periods $\leq50$ minutes. }%
\label{FIG_iono_tec_athens_2017_sept_7-10_storm_5min__50period}%
\end{center}
\end{figure}

We see that during the geomagnetic storm we have a lot of wave-packages and
regular patterns. However what is not less interesting, similar pattern appear
during the two days $9$ and $10$ September, (during the recovery phase of the
storm), at midday time, having periods $20-50$ minutes. This seems to be due
to the solar terminator influence, as was suggested in \cite{Belashov2015},
\cite{Schnepf2016}.

\subsubsection{Experiments with IMF data from ACE satellite, 7-10 Sept., 2017,
$4$ min. data}

We retrieved the $B_{z}$ component of the IMF from the ACE satellite on 7-10
Sept., 2017, every 4 min. data.

\begin{figure}
[!ptb]
\begin{center}
\includegraphics[
height=3.0in,
width=5.0in
]%
{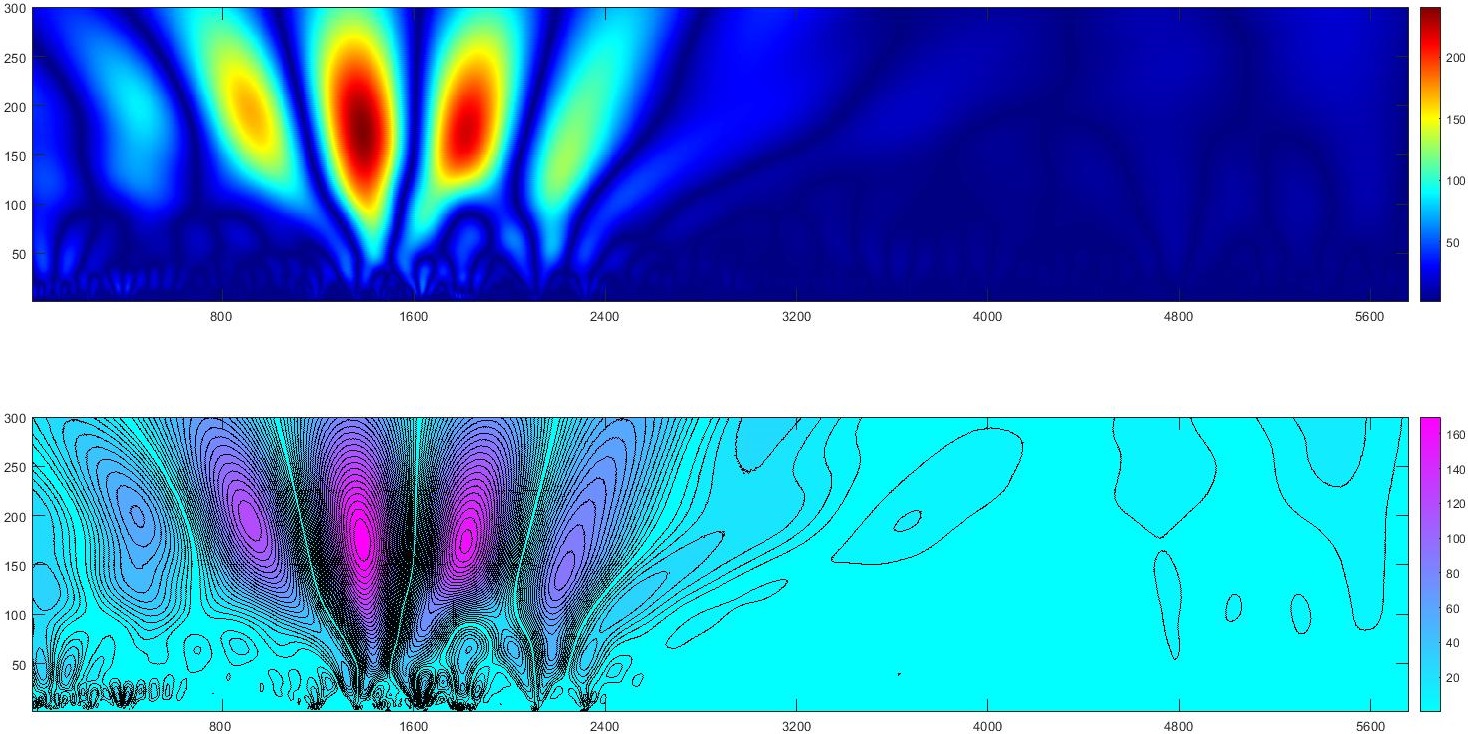}%
\caption{\textbf{Heatmap} is on the top and the \textbf{contour} \textbf{map}
is on the bottom, for CWT of $B_{z}$ component of IMF, $7-10$ Sept., $2017.$ }%
\label{FIG_interplan_ace_bz__2017_sept_7-10_storm_4min}%
\end{center}
\end{figure}

On Figure \ref{FIG_interplan_ace_bz__2017_sept_7-10_storm_4min} we provide the
CWT of the time series containing the values of $B_{z}$; again the
\textbf{heatmap} is on the top and the \textbf{contour} \textbf{map} is on the
bottom of the Figure.%

An interesting observation is that in the ACE data and in the ground PAG data
one \textbf{cannot identify} any wave-packages of periods $20-100$ min, after
the storm, i.e. on Sept. $9,10,$ $2017.$ This makes us believe that there is a
correlation between the two observable values. This is in a strong contrast to
the Ionospheric observations provided above on Figure
\ref{FIG_iono_tec_athens_2017_sept_7-10_storm_5min} and Figure
\ref{FIG_iono_tec_athens_2017_sept_7-10_storm_5min__50period}, where such wave
packages are available. This shows that in some cases the ionospheric plasma
generates short-period modes with a relatively small amplitude. This would
explain the lack of similar modes in the ground geomagnetic data of PAG. On
the other hand, it is clear that on $9$ and $10$ September, these
short-periodic modes are the result of eigen-oscillations of the ionospheric
plasma, which are not caused by the influence of the Interplanetary Magnetic field.

\subsection{Experiments with data for the 2003 strong geomagnetic storm}

Since the geomagnetic storm in 2003 was unusually strong, it has become as a
handbook example for the testing the analysis tools. However, in 2003 there
were not so many data available. In particular, the \textbf{ionosound} data
are not available with the present sampling, but only hour data.

We provide below the results for Wavelet Analysis for the following data

1.\qquad In $2003$, for different (ground) geomagnetic ground observatories
(SUA) we have 1 minute data for the \ $H-$component.

2.\qquad ACE satellite data for IMF are available at frequency $4$ min.

On the other hand, during this storm the Ionospheric data are only
\textbf{mean-hour} which is \textbf{insufficient} for the present analysis.

First of all, we have the main trend of the magnetic fields provided by the
$D_{st}$ data on Figure \ref{FIG_dst1_kyoto} above.

\subsubsection{Experiments with geomagnetic data from SUA, on 28-29 Oct.,
2003}

One may analyze the original source data $H-$component, or $DS$ obtained after
subtracting the $D_{st}$ from the data, see formula (\ref{D=Dst-Defined}). As
we said above, since the $D_{st}$ data are given every hour this may create
artefacts every whole hour. We provide the contour plot of the CWT for the
$H-$ component of the geomagnetic field from SUA, 28-29 Oct., 2003, sampling 1
minute, on the following Figure
\ref{FIG_ds_1_wavelet_sua_min_storm1_sym8_2048___contour__2__no_dst_removed}.%

\begin{figure}
[!ptb]
\begin{center}
\includegraphics[
height=3.0in,
width=5.0in
]%
{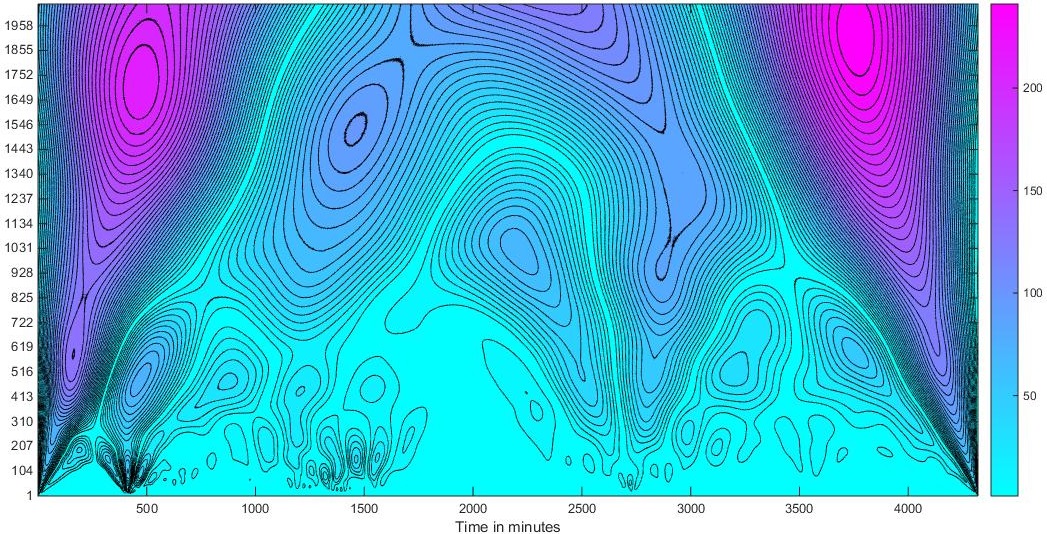}%
\caption{Contour plot of CWT for the $H-$ component from SUA, $28-29$ Oct.,
$2003.$ }%
\label{FIG_ds_1_wavelet_sua_min_storm1_sym8_2048___contour__2__no_dst_removed}%
\end{center}
\end{figure}

On Figure
\ref{FIG_ds_1_wavelet_sua_min_storm1_sym8_2048___contour__2__no_dst_removed},
we may clearly identify short-period wave packages with periods below $3-4$
hours, as well as with periods $4-12$ hours, but also with periods about $24$
hours. The last are related with the main and recovery phase of the storm,
i.e. with the Ring current. The wave packages with periods below $3$ hours may
ber related to the fluctuations of the Ring current and the ionospheric
instabilities generating such modes. Modes with periods below $100$ minutes
are generated mainly at midday, and may be related to the solar terminatory
and are eventually of soliton type, as was mentioned already above, see also
\cite{Belashov2015}, \cite{SrebrovPashovaKounchev}.

\begin{figure}
[!ptb]
\begin{center}
\includegraphics[
height=3.0in,
width=5.0in
]%
{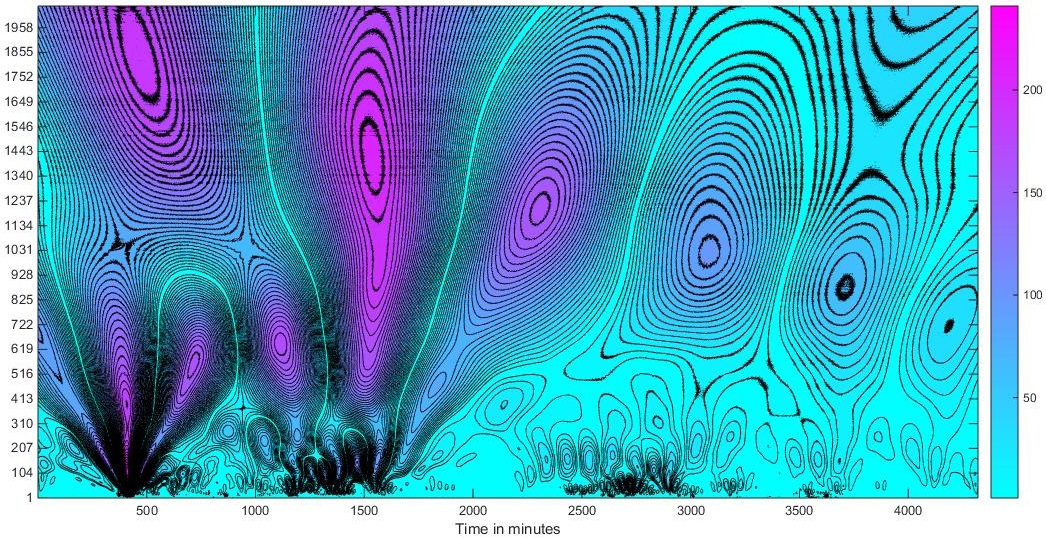}%
\caption{contour plot of the CWT for the $DS$ index of the geomagnetic field
from SUA, $28-29$ Oct., $2003.$}%
\label{FIG_ds_1_wavelet_sua_min_storm1_sym8_2048___contour.}%
\end{center}
\end{figure}

On Figure \ref{FIG_ds_1_wavelet_sua_min_storm1_sym8_2048___contour.} we
provide the CWT of the $DS$ data.%

\subsubsection{Experiments with spline smoothing of $D_{st}$}

\begin{figure}
[!ptb]
\begin{center}
\includegraphics[
height=3.0in,
width=5.0in
]%
{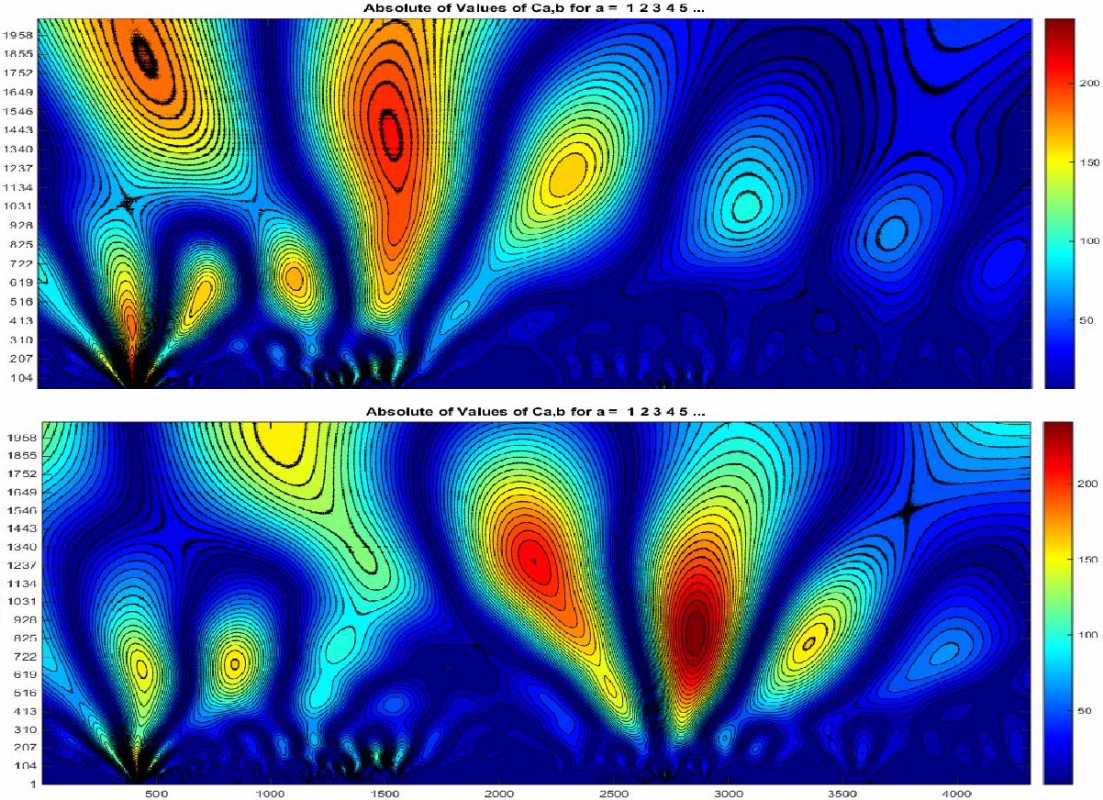}%
\caption{...}
\label{Fig_SplineDst}%
\end{center}
\end{figure}

We have provided some interesting experiments which show the effect of
subtracting of $D_{st}$ after smoothing the $D_{st}$ with splines, Figure
\ref{Fig_SplineDst}.%

On the bottom of Figure \ref{Fig_SplineDst} we have the CWT of the
$H-$component and on the top we have the CWT of the $DS=H-\widetilde{D_{st}}$
where $\widetilde{D_{st}}$ is the smoothed $D_{st}$ (which as mentioned is
provided by the WDC of geomagnetism in Kyoto on hourly basis).

This shows that one has to be careful when subtracting the $D_{st}$ index
(which is a step function) from the $H-$component since this creates
non-smooth signal and the Fourier or Wavelet analyses generate artificial frequencies.

\subsubsection{Experments with IMF data, on 28-29 Oct., 2003, $4$ min. data}

We have provided the CWT for the $B_{z}$ component of the IMF on the following
Figure \ref{FIG_interplan_mag_28-29oct_2013__4min_data__z_component}. Again
the \textbf{heatmap} of the CWT is on the top, while the \textbf{contour} map
is on the bottom.%

\begin{figure}
[!ptb]
\begin{center}
\includegraphics[
height=3.0in,
width=5.0in
]%
{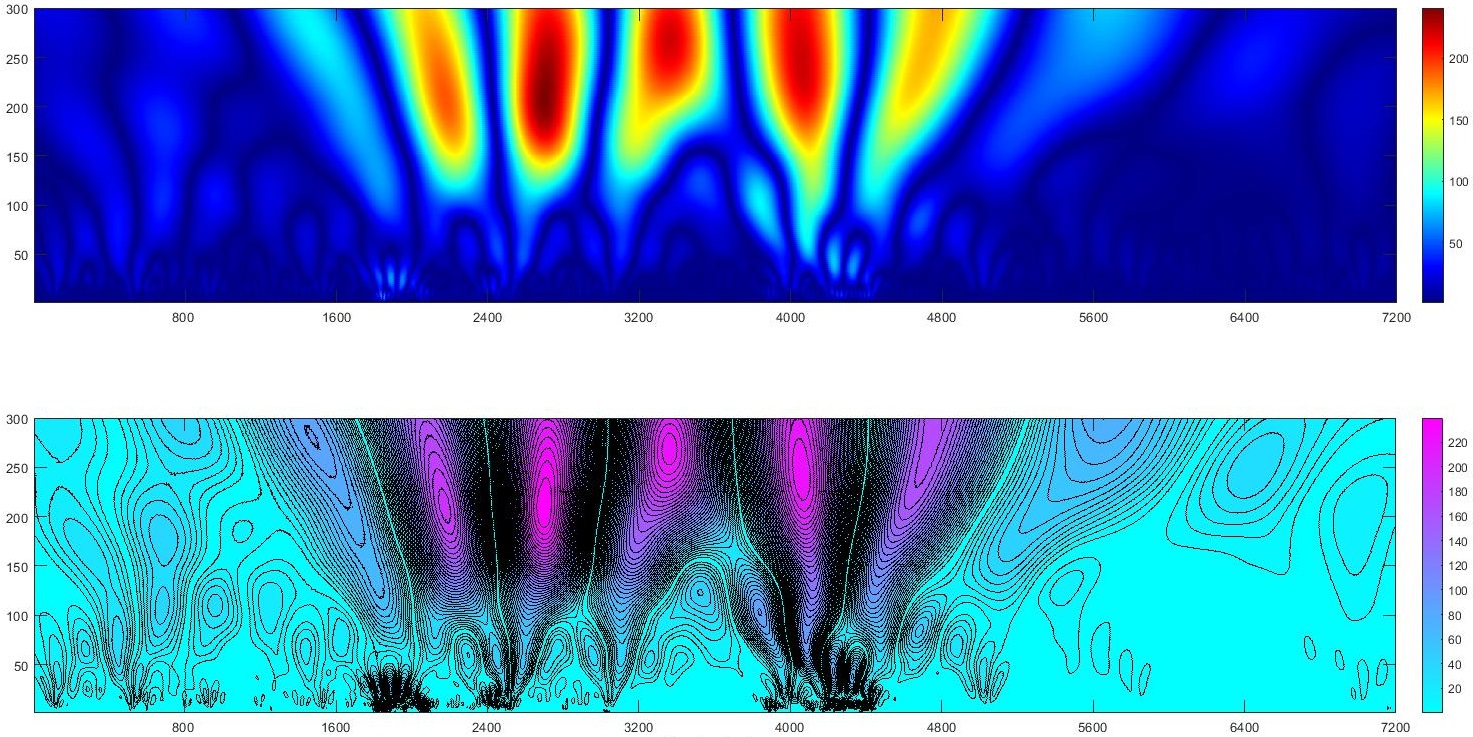}%
\caption{\textbf{Heatmap} of the CWT on the top, and the \textbf{contour} map
on the bottom, \textbf{for the }$B_{z}$ component of IMF, $28-29$ Oct.,
$2003.$ }%
\label{FIG_interplan_mag_28-29oct_2013__4min_data__z_component}%
\end{center}
\end{figure}

On the Figure \ref{FIG_interplan_mag_28-29oct_2013__4min_data__z_component} we
see various families of wave packages, which may be separated into two
types:\ those with periods less than $100$ minutes, and those with periods
between $100-450$ minutes. Their explanation is related to the complex
structure of the disturbance of the $B_{z}$ component during the strong
geomagnetic storm.

\section{CONCLUSIONS}

\begin{enumerate}
\item The main objective of the present research is to apply Wavelet Analysis
to Big data in the Solar-Terrestrial Physics, for the investigation of short
period variations of the (ground) \textbf{geomagnetic field/Ionospheric
parameters} in a region with mean geographic latitude. Thus, by applying
Continuous Wavelet Transform to large amount of heterogeneous data
(geomagnetic field, ionospheric parameters and IMF), we have identified modes
(wave packages) with different periods, of the order of $20$ to few hundred
minutes with a significant amplitude, which is enough to be registered by the
equipment in the geomagnetic observatories.

As it is known, in the same range there exist the so-called geomagnetic
pulsations, but they have a very low amplitude and exist for a short time only
during the night hours for this geographic latitude. Unlike the geomagnetic
pulsations, the short period variations of our interest, have significant
amplitude, and are identified in the present research; they are discovered
during the whole day and may be divided into modes with periods less than $3$
hours and modes which have a period greater than $3$ hours.

\item Our analysis of the variations of the geomagnetic field, the ionospheric
plasma parameters, and the IMF, has shown persistent short term periodic
events, as wave packages. The short period modes (wave packages) of the
variations which we have identified have a clear explanation (e.g. from plasma
physics) and are caused by macro-instabilities in different domains of the
near Earth space environment.

\item We have identified the presence of modes with periods lower than $3$
hours, generated predominantly by the ionospheric plasma, but also similar
modes which exist in the IMF. The ability of Wavelet Analysis to uncover
Multiresolution structure of the data, gave us possibility to identify
short-periodic wave packages in the geomagnetic field variations, in IMF and
in the Ionospheric parameters.

\item The present research represents a contribution to the newly developing
area of AstroGeoInformatics due to the large spectrum of the analyzed
phenomena which belong to the Solar-Terrestrial Physics.
\end{enumerate}

\section{Thanks}

The authors due thanks to the editors of this volume, Petr Skoda and Adam
Fathalrahman, for the patience and their assistance. Thanks extend to
Aleksandra Nina (Belgrade) for the discussions on the ionosphere. The two
first-named authors thank the Project on Modern mathematical methods for Big
Data, DH 02-13 with Bulgarian NSF, and also the Project SatWebMare with ESA
(in the PECS framework). Last but not least, all owe thanks to the COST action
BigSkyEarth with EU. OK thanks the Alexander von Humboldt Foundation.

The services of the World Data Center for Geomagnetism, at the Kyoto
University, Japan, the INTERMAGNET network, the ACE Science Center at Caltech,
and the GIRO center, University of Massachusettes at Lowell, are gratefully acknowledged.

\section{APPENDIX on Wavelet Analysis and its applications to geomagnetic
data\label{AppendixWavelets}}

In the present research we have decided for Continous Wavelet Transform (CWT).
We provide the essentials of the CWT and some useful references for the
applications of Wavelet Analysis.

\subsection{Technical stuff}

We will say that the integrable function $\psi$ is a \textbf{wavelet function}
if it satisfies the following properties:\ 

\begin{enumerate}
\item the \textbf{admissibility} condition holds%
\[
0<C_{\psi}:=\int_{-\infty}^{\infty}\frac{\left\vert \widehat{\psi}\left(
\omega\right)  \right\vert ^{2}}{\left\vert \omega\right\vert }d\omega<\infty
\]

\item the \textbf{zero integral} condition holds
\[
\widehat{\psi}\left(  0\right)  =0
\]
where $\widehat{\psi}\left(  \omega\right)  $ is the Fourier transform of
$\psi.$ This condition is equivalent to
\[
\int\psi\left(  \omega\right)  d\omega=0.
\]

\end{enumerate}

We consider only real valued functions $\psi.$

Once the wavelet function $\psi$ is fixed, then for every integrable function
$f$ (which is considered to represent the signal) which has a sufficient decay
at $\infty,$ and for every two real numbers $a,b\in\mathbb{R}$ with $a>0,$ we
may define the CWT $W_{\psi}f\left(  a,b\right)  $ by putting:\
\begin{equation}
W_{\psi}f\left(  a,b\right)  :=\frac{1}{\sqrt{a}}\int_{-\infty}^{\infty
}f\left(  t\right)  \psi\left(  \frac{t-b}{a}\right)  dt \label{CWTdefinition}%
\end{equation}
The number $a$ is called \textbf{scale}, and $b$ is called
\textbf{translation} (shift). Recall that the usual definition of the
\textbf{frequency} $k$ is then given by putting
\[
k=\frac{1}{a}.
\]

Unlike the usual Fourier transform where the dimension of the variable of the
signal $f\left(  t\right)  $ is transformed into the same dimensional
frequency domain, here we see that the CWT $W_{\psi}f\left(  a,b\right)  $
depends on two variables. We are able to reconstruct the original signal $f$
from this representation, by means of the \textbf{Calderon} inversion formula
(\cite{jaffard}, \cite{mallat}):\
\begin{equation}
f\left(  t\right)  =\frac{1}{C_{\psi}}%
{\displaystyle\int_{-\infty}^{\infty}}
{\displaystyle\int_{-\infty}^{\infty}}
\frac{1}{\sqrt{a}}W_{\psi}f\left(  a,b\right)  \psi\left(  \frac{t-b}%
{a}\right)  \frac{da}{a^{2}}db \label{Calderon}%
\end{equation}

As in the DWT it is always the question to find some reasonable approximation
in the Calderon formula which takes into account only the larger values of
$\left\vert W_{\psi}f\left(  a,b\right)  \right\vert $, which will result in
an approximation of the double integral in the equality in formula
(\ref{Calderon}). Thus, the question is, whether it is possible to use just a
part of the integration domain? This may be achieved in different ways; one
approach is to apply a a threshold on the absolute value of the CWT
$\left\vert W_{\psi}f\left(  a,b\right)  \right\vert $, say $\varepsilon>0,$
and define the domain
\[
D_{\varepsilon}=\left\{  \left(  a,b\right)  :\left\vert W_{\psi}f\left(
a,b\right)  \right\vert <\varepsilon\right\}
\]
and then consider the approximation integral
\[
I_{\varepsilon}:=\frac{1}{C_{\psi}}\underset{D_{\varepsilon}}{%
{\displaystyle\int}
{\displaystyle\int}
}\frac{1}{\sqrt{a}}W_{\psi}f\left(  a,b\right)  \psi\left(  \frac{t-b}%
{a}\right)  \frac{da}{a^{2}}db.
\]
so that the remainder would satisfy
\[
\left\vert f\left(  t\right)  -I_{\varepsilon}\left(  t\right)  \right\vert
\leq\delta\qquad\text{for all }t\in\mathbb{R}%
\]

Unlike the discrete wavelet transform (DWT) here we do not have a clearly
defined Multi-resolution Analysis (MRA), and also there are no father wavelets
(scaling functions). However, in general, one may use the wavelets in the
Discrete Wavelet theory and apply them in CWT if they are smooth enough. The
CWT is very convenient tool to detect and characterize singularities in
functions, in order to distinguish between noise and signal (see
\cite{jaffard}, \cite{mallat}). In particular, one may use CWT to study
fractal behaviour of the signals.

\begin{figure}
[!ptb]
\begin{center}
\includegraphics[
height=3.0448in,
width=3.6629in
]%
{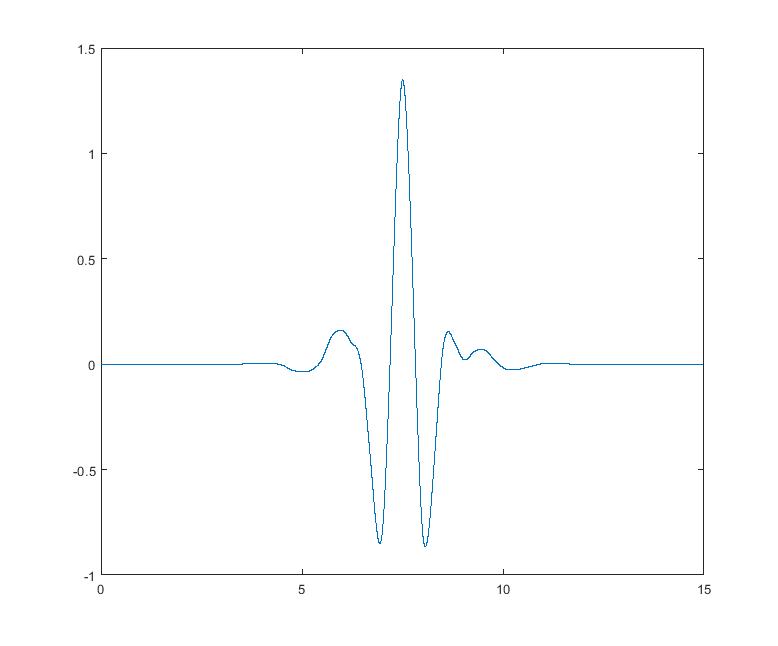}%
\caption{The graph of the sym8 mother wavelet. }%
\label{Sym8}%
\end{center}
\end{figure}

In a wide area of applications, people use CWT with a wavelet function $\psi$
equal to the Mexican hat and the Morlet wavelet, although these two functions
do not enjoy the usual scheme of Multiresolution Analysis as introduced in
\cite{mallat}, \cite{jaffard}. In the present research, after numerous
experimentations, we have decided for the \textbf{symlet} family of functions,
which are a modified version of Daubechies wavelets family \textbf{db,} since
they enjoy increased symmetry, \cite{addison}. We have applied the
\textbf{sym8} wavelet function, provided on Figure \ref{Sym8} below (see also
the website http://wavelets.pybytes.com/wavelet/sym8/).%

However it is important to remark that the experiments with many other
wavelets $\psi,$ have shown that the singularities which we detect by the
\textbf{symlets} may be analyzed with the same succes by applying the other
wavelets; completely subjectively, we have found that \textbf{sym8} gives in
average one of the best possible visual picture. This fact shows that our
observation is due to persistent physical events and may not be an artefact
which is due to the particular wavelet which we choose.

\subsection{CWT of some simple functions \label{sectionImpulseTrain}}

An important control of the Wavelet Analysis method is to consider the CWT of
the simple jump functions, as for example \textbf{impulse trains} which are
sums of Dirac delta functions. It gives us idea about the behavior of the CWT
of more complicated signals. A main reason to consider these impulse trains is
the result of Belashov \cite{Belashov2015}, who has very successfully modeled
(better than the traditional IRI model) the ionospheric data by assuming that
the disturbances of the Ionosphere are related (even in quiet days!) to wave
packages having soliton character; we have already announced this resemblance
of our analysis in \cite{SrebrovPashovaKounchev}.

The following Figure \ref{Impulse} shows the graph of a simple impulse train.%

\begin{figure}
[!ptb]
\begin{center}
\includegraphics[
height=1.5287in,
width=3.2316in
]%
{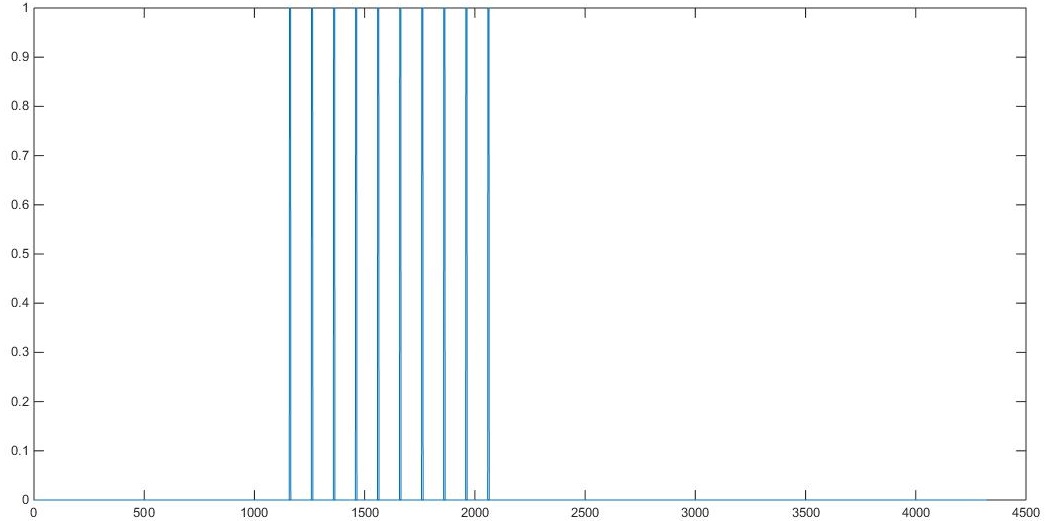}%
\caption{Impulse train}%
\label{Impulse}%
\end{center}
\end{figure}

\begin{figure}
[!ptb]
\begin{center}
\includegraphics[
height=1.7299in,
width=3.6566in
]%
{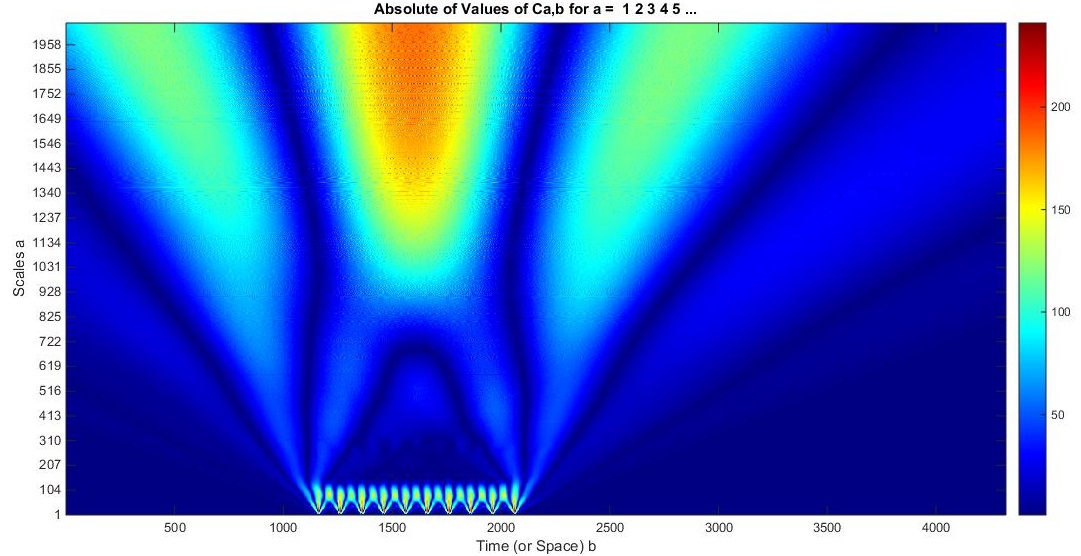}%
\caption{CWT of the impulse train. }%
\label{ImpulseCWT}%
\end{center}
\end{figure}

It has the CWT shown on Figure \ref{ImpulseCWT}. The $y-$axis shows the lenght
of the period (the scale $a$ in the CWT $W_{\psi}\left(  a,b\right)  $ and the
$x-$axis shows the number $b$).%

\begin{remark}
From the above Figure \ref{ImpulseCWT} we see that a package of periodic
pulses has considerable CWT $\left\vert W_{\psi}\left(  a,b\right)
\right\vert $ for lower periods $0\leq a\leq105,$ but it also shows an
\textquotedblleft integral effect\textquotedblright\ and shows a considerable
CWT $\left\vert W_{\psi}\left(  a,b\right)  \right\vert $ for longer periods
$a\geq1030.$ The moral of this observation is that one has to be really
careful when solving the Inverse problem, i.e. when making conclusion about
the singularities of the original signal $f\left(  t\right)  $ judging by the
large period behaviour of the CWT $W_{\psi}\left(  a,b\right)  .$
\end{remark}

\begin{remark}
Another interesting example is provided also in Wikipedia,

https://en.wikipedia.org/wiki/Continuous\_wavelet\_transform, where CWT is
provided of a frequency breakdown signal by using the \textbf{symlet} as a
wavelet function with 5 vanishing moments, see section \ref{AppendixWavelets}.
\end{remark}


\begin{thebibliography}{99}                                                                                               %


\bibitem {addison}Addison, $2011.$, \textit{The Illustrated Wavelet Transform
Handbook: Introductory Theory and Applications in Science, Engineering,
Medicine and Finance, Second Edition}, CRC Press

\bibitem {akasofu}Akasofu, S., S. Chapman, $1972.$, \textit{Solar-Terrestrial
Physics}, Oxford at the Clarendon Press.

\bibitem {artigas}M Zossi de Artigas, P Fernandez de Campra, E. M. Zotto,
jul./sep. $2008.$, \textit{Geomagnetic disturbances analysis using discrete
wavelets}, Geof\'{\i}s. Intl vol.47 no.3 M\'{e}xico

\bibitem {Belashov2015}Belashov, V. Yu, E. S.Belashova, $2015.$,
\textit{Dynamics of IGW and traveling ionospheric disturbances in regions with
sharp gradients of the ionospheric parameters}, Advances in Space Research,
Vol. 56, Issue 2, 333-340.

\bibitem {bishop}Bishop, C., $2006.$, \textit{Pattern Recognition and Machine
Learning}, Springer, New York

\bibitem {boudouris}A. Boudouridis and E. Zesta, $2007.$, \textit{Comparison
of Fourier and wavelet techniques in the determination of geomagnetic field
line resonances}, J. of Geoph. Res., vol. 112, A08205.

\bibitem {Braitenberg}Carla Braitenberg, Alexander B. Rabinovich (eds.),
$2017.$, \textit{The Chile-2015 (Illapel) Earthquake and Tsunami},
Birkh\"{a}user Basel

\bibitem {frick}Frick P, Baliunas SL, Galyagin D et al ($1997.$),
\textit{Wavelet analysis of stellar chromospheric activity variations},
Astrophys J 483:426--434.

\bibitem {gencay}Gen\c{c}ay, R., Sel\c{c}uk, F., Whitcher, B., $2002.$,
\textit{An Introduction to Wavelets and Other Filtering Methods in Finance and
Economics}, Academic Press, San Diego

\bibitem {Heisler}Heisler, L.H., $1959.$, \textit{Occurrence of giant
travelling ionospheric disturbances at night}, Nature 183, 383--384.

\bibitem {Hocke1996}Hocke, K., and K. Schlegel, $1996.$, \textit{A review of
atmospheric gravity waves and traveling ionospheric disturbances: 1982--1995},
Ann. Geo-phys.,14, 917--940

\bibitem {hubbard}Barbara Burke Hubbard, $1998.$, \textit{The World According
to Wavelets: The Story of a Mathematical Technique in the Making}, AK Peters Ltd

\bibitem {Hunsucker1987}Hunsucker, R.D., $1987.$, \textit{The sources of
gravity waves}, Nature 328, 204--205.

\bibitem {Jach}A. Jach, P. Kokoszka, J. Sojka, and L. Zhu, $2006.$,
\textit{Wavelet-based index of magnetic storm activity}, J. of Geoph. Res.,
vol. 111, A09215.

\bibitem {jaffard}S. Jaffard, Y. Meyer, and R.D. Ryan, $2001.$,
\textit{Wavelets. Tools for science \& technology}, SIAM, Philadelphia

\bibitem {katsavrias}Ch. Katsavrias, A. Hillaris, P. Preka-Papadema, $2016.$,
\textit{A Wavelet Based Approach to Solar--Terrestrial Coupling}, Advances in
Space Research, Volume 57, Issue 10, 15 May 2016, Pages 2234-2244.

\bibitem {Klausner2011}V. Klausner, M. O. Domingues, O. Mendes, A. R.R. Papa,
$2011.$, \textit{Tsunami effects on the Z component of the geomagnetic field}, https://arxiv.org/abs/1108.4893

\bibitem {Klausner2014}V. Klausner, O Mendes, MO Domingues, ARR Papa, RH
Tyler, P Frick, Esfhan A Kherani, $2014.$, \textit{Advantage of wavelet
technique to highlight the observed geomagnetic perturbations linked to the
Chilean tsunami (2010)}, Journal of Geophysical Research: Space Physics 119
(4), 3077-3093.

\bibitem {Klausner2016}Virginia Klausner, Esfhan A. Kherani, and Marcio T. A.
H. Muella, February $2016.$, \textit{Near- and far-field tsunamigenic effects
on the Z component of the geomagnetic field during the Japanese event}, J.
Geoph. Research, Space Physics

\bibitem {Klausner2017}V. Klausner, T. Almeida, F. C. De Meneses, E. A.
Kherani, V. G. Pillat and M. T. A. H. Muella, ($2017.$), \textit{Chile2015:
Induced Magnetic Fields on the Z Component by Tsunami Wave Propagation, The
Chile-2015 (Illapel) Earthquake and Tsunami}, 10.1007/978-3-319-57822-4\_14, (193-208)

\bibitem {Klausner2016b}V. Klausner, T. Almeida, F. C. de Meneses, E. A.
Kherani, V. G. Pillat and M. T. A. H. Muella, ($2016.$), \textit{Chile2015:
Induced Magnetic Fields on the Z Component by Tsunami Wave Propagation}, Pure
and Applied Geophysics, 173, 5, (1463)

\bibitem {kounchevBOOK}O. Kounchev, $2001.$, \textit{Multivariate Polysplines.
Applications to Numerical and Wavelet Analysis}, Academic Press/Elsevier, San Diego

\bibitem {Leimohn}Liemohn, M. W. ($2003.$), \textit{Yet another caveat to the
Dessler-Parker-Sckopke relation}, J. Geophys. Res., 108(A6), 1251.

\bibitem {mallat}Mallat, S., $2009.$, \textit{A Wavelet Tour of Signal
Processing}, Academic Press, Burlington

\bibitem {Mandrikova}Mandrikova, O. V., I. S. Solovev, and T. L Zalyaev,
$2014.$, \textit{Methods of analysis of geomagnetic field variations and
cosmic ray data}, Earth, Planets and Space, 66:148.

\bibitem {MeinlSun}Meinl, T., Sun, E., $2012.$, \textit{A nonlinear filtering
algorithm based on wavelet transforms for high-frequency financial data
analysis}, Stud.Nonlinear Dyn. E 16(3), 5.

\bibitem {Mitchner}M. Mitchner, Charles H. Kruger, $1973$, \textit{Partially
Ionized Gases}, (Wiley series in plasma physics)

\bibitem {mullon}L. Mullon, N. B. Chang, Y. J. Yang, and J. Weiss, ($2013.$),
\textit{Integrated remote sensing and wavelet analyses for screening short
term teleconnection patterns in northeast America}, Journal of Hydrology 499: 247--264.

\bibitem {oppenheim}Oppenheim, A., Schafer, R., $2010.$, \textit{Discrete-Time
Signal Processing, 3rd edition}, Prentice-Hall, Upper Saddle River

\bibitem {percival}Donald B. Percival and Andrew T. Walden, $2000.$,
\textit{Wavelet Methods for Time Series Analysis}, Cambridge University Press

\bibitem {Schnepf2016}Schnepf NR, Manoj C, An C et al., ($2016.$),
\textit{Time-frequency characteristics of tsunami magnetic signals from four
Pacific Ocean events}, Pure appl Geophys 173:3935--3953.

\bibitem {srebrov2003}B.A. Srebrov, $2003.$, \textit{MHD Modeling of
supersonic, super-alfvenic distrubances propagating in the interplanetary
plasma and their relationship to the geospace environment}, \ Advances in
Space Research, Volume 31, Issue 5, March 2003, Pages 1413-1418.

\bibitem {SrebrovPashovaKounchev}B. Srebrov, L. Pashova, O. Kounchev, $2018.$,
\textit{Study of Local Manifestations of G5 -- Extreme Geomagnetic Storms
(29--31 October, 2003) in Midlatitudes Using Geomagnetic Data by Continuous
Wavelet Transforms}, C.R. acad. bulg. sci., Volume 71, Issue No6, 2018.

\bibitem {Sun2012}Edward W. Sun, Thomas Meinl, $2012.$, \textit{A new
wavelet-based denoising algorithm for high-frequency financial data mining},
European Journal of Operational Research, Volume 217, Issue 3, 16 March 2012,
Pages 589-599

\bibitem {Sun2015}Edward W. Sun, Min-The Yu, $2015.$, \textit{Generalized
optimal wavelet decomposing algorithm for big financial data}, International
Journal of Production Economics Volume 165, July 2015, Pages 194-214

\bibitem {Wei}Wei, H. L, S. A. Billings, and M. Balikhin, ($2004.$),
\textit{Analysis of the geomagnetic activity of the Dst index and self-affine
fractals using wavelet transforms}, Nonlinear Processes in Geophysics (2004)
11: 303--312

\bibitem {xu}Z. Xu, L. Zhu, J. J. Sojka, P. Kokoszka, and A. Jach, $2008.$,
\textit{An assessment study of the wavelet-based index of magnetic storm
activity (WISA) and its comparison to the dst},\ J. Geophys. Res., vol. 70,
pp. 1579-1588

\bibitem {xu2011}Z. Xu, $2011.$, \textit{Study of Geomagnetic Disturbances and
Ring Current Variability During Storm and Quiet Times Using Wavelet Analysis
and Ground-based Magnetic Data from Multiple Stations}, PhD thesis, Utah State University
\end{thebibliography}
\end{document}